\documentclass[trackchanges]{aastex701}

\newcommand{\CO}{\ensuremath{\rm ^{12}CO}}
\newcommand{\COO}{\ensuremath{\rm ^{13}CO}}
\newcommand{\COOO}{\ensuremath{\rm C^{18}O}}
\newcommand{\EBV}{\ensuremath{E(B-V)}}

\newcommand{\vlsr}{\ensuremath{\,{\rm V_{LSR}}}}
\newcommand{\mum}{\ensuremath{\,{\rm \mu m}}}

\newcommand{\kpc}{\ensuremath{\,{\rm kpc}}}
\newcommand{\K}{\ensuremath{\,{\rm K}}}

\newcommand{\kkms}{\ensuremath{\,{\rm K~ km\, s^{-1}}}}
\newcommand{\kms}{\ensuremath{\,{\rm km\, s^{-1}}}}
\newcommand{\per}{\ensuremath{\,{\rm \%}}}

\usepackage{booktabs}
\usepackage{tablefootnote}
\usepackage{subfigure}
\usepackage{graphicx}
\usepackage[utf8]{inputenc}
\usepackage{CJK}
\usepackage{amsmath}
\usepackage{makecell}
\usepackage{soul}
\usepackage{bm}

\begin{document}
\title{The Extinction Distance of Supernova Remnants in Combination with the CO Line Measurements}

\begin{CJK*}{UTF8}{gbsn}
\correspondingauthor{Jun Li, Biwei Jiang}
\email{lijun@gzhu.edu.cn, bjiang@bnu.edu.cn}

\author[0000-0002-1007-3700]{Zhe Zhang (张哲)}
\affiliation{School of Physics and Astronomy, Beijing Normal University, Beijing 100875, People's Republic of China}
\affiliation{Institute for Frontiers in Astronomy and Astrophysics, Beijing Normal University, Beijing 102206, People's Republic of China}
\email{zhangz324@mail.bnu.edu.cn}

\author[0000-0001-9328-4302]{Jun Li (李军)}
\affiliation{Center for Astrophysics, Guangzhou University, Guangzhou 510006, People's Republic of China}
\email{lijun@gzhu.edu.cn}

\author[0000-0003-3168-2617]{Biwei Jiang (姜碧沩)}
\affiliation{School of Physics and Astronomy, Beijing Normal University, Beijing 100875, People's Republic of China}
\affiliation{Institute for Frontiers in Astronomy and Astrophysics, Beijing Normal University, Beijing 102206, People's Republic of China}
\email{bjiang@bnu.edu.cn}

\author[0000-0003-2645-6869]{He Zhao (赵赫)}
\affiliation{Departamento de Fisica y Astronomia, Facultad de Ciencias Exactas, Universidad Andres Bello, Fernandez Concha 700, 8320000 Santiago, Chile}
\email{he.zhao@oca.eu}

\author[0009-0007-3238-4212]{Fupeng Liu (刘付堋)}
\affiliation{School of Physics and Astronomy, Beijing Normal University, Beijing 100875, People's Republic of China}
\affiliation{Institute for Frontiers in Astronomy and Astrophysics, Beijing Normal University, Beijing 102206, People's Republic of China}
\email{202421101131@mail.bnu.edu.cn}


\begin{abstract}
Accurate distance measurements to supernova remnants (SNRs) are crucial for understanding their physical properties and evolution. We present a novel method that combines CO line observations with three-dimensional (3D) extinction maps to determine distances to SNRs (G93.7$-$0.2, G109.1$-$1.0, G156.2+5.7, and G166.0+4.3) through their associated molecular clouds. For each SNR, candidate CO velocity components corresponding to interacting molecular clouds are identified based on previous observational evidence with refinements: [$-19, -3$] $\kms$ for G93.7$-$0.2, [$-51, -46$] $\kms$ for G109.1$-$1.0, [$-$10, 0] $\kms$ for G156.2+5.7, and [$-27, -15$] $\kms$ for G166.0+4.3. By examining extinction-distance profiles along the sightlines and identifying extinction jumps that spatially coincide with CO emission features, we derive distances of $1.82 \pm 0.13$ kpc for G93.7$-$0.2, $3.05 \pm 0.15$ kpc for G109.1$-$1.0, $0.60 \pm 0.15$ kpc for G156.2+5.7, and $3.44 \pm 0.23$ kpc for G166.0+4.3. Our extinction-based distances are largely consistent with previous estimates while with better accuracy and robustness. 

\end{abstract}


\section{Introduction} \label{sec:intro}
Establishing a reliable distance to a supernova remnant (SNR) is the cornerstone for interpreting its observational characteristics and evolutionary history, and is essential for deriving SNR's age, linear size, and explosion energy, which ultimately reveal the nature of the progenitor and the underlying explosion mechanism \citep{Ranasinghe2018MNRAS}. Despite decades of dedicated research, persistent distance uncertainties remain a primary bottleneck in SNR studies \citep{Zhu2014IAUS,Chen2025ApJ}.

Various methods have been developed to measure SNR distances, each with inherent advantages and limitations. The kinematic distance method, one of the most widely employed approaches, utilizes H\MakeUppercase{\romannumeral1} absorption spectra or CO emission spectra from the clouds associated with SNRs \citep{Tian2007A&A,Ranasinghe2017ApJ}. Systemic velocities derived from these spectral lines are converted to distances using Galactic rotation curve models \citep{Roman-Duval2009ApJ,Reid2016ApJ}. However, this method suffers from the well-known kinematic distance ambiguity in the inner Galaxy, where a single radial velocity corresponds to two possible distances \citep{Roman-Duval2009ApJ}. Additionally, deviations from circular rotation caused by spiral arm perturbations and streaming motions can introduce systematic uncertainties in the derived distances \citep{Gomez2006AJ, Xu2006Sci}. The empirical $\Sigma$--$D$ relation, which correlates the radio surface brightness ($\Sigma$) of shell-type SNRs with their physical diameter ($D$) through a power-law relationship ($\Sigma_{\nu} \propto D^{-\beta}$), provides a statistical approach to distance estimation \citep{Case1998ApJ,Guseinov2003A&AT,Pavlovic2013ApJS}. While useful for population studies, this method yields significant uncertainties for individual objects due to intrinsic variations in SNR evolution, environmental effects, and selection biases \citep{Urosevic2005A&A,Pavlovic2018ApJ}. Furthermore, the comparison of distances derived from different $\Sigma$--$D$ relations to those derived from other reliable methods shows a discrepancy of 50$\per$ on average \citep{Ranasinghe2023ApJS}.

More precise but less generally applicable methods exist for specific SNRs. Direct parallax measurements of associated pulsars using very long baseline interferometry (VLBI) can provide distance accuracies better than 10$\per$ \citep{Dodson2003ApJ}. Proper motion measurements of X-ray filaments, combined with shock velocity estimates from spectroscopy, offer another geometric distance technique \citep{Katsuda2008ApJ}. However, these methods require either the presence of a detectable pulsar or high-resolution X-ray observations spanning many years, limiting their applicability to a small fraction of the known SNR population.

Three-dimensional (3D) dust extinction mapping has emerged as a powerful tool for distance determination, offering a reliable approach that avoids the kinematic distance ambiguity \citep{Stead2010MNRAS}. Two primary strategies have been employed to derive SNR distances from extinction data. The first uses red clump stars (RCs) as standard candles to establish distance-extinction relationships toward SNRs with known extinction values \citep{Shan2018ApJS}. The second identifies extinction increases (``jumps") along sightlines toward SNRs by analyzing stellar extinction-distance profiles \citep{Zhao2018ApJ,Zhao2020ApJ,Wang2020AA,Chen2025ApJ}. However, since most SNRs reside in the dusty Galactic plane where molecular clouds (MCs) dominate the extinction signal, these jumps often trace foreground or background MCs rather than the SNRs themselves. Without independent evidence linking an SNR to a specific MC, extinction jumps alone cannot reliably constrain SNR distances. 

Observational evidence indicates that many SNRs interact with nearby MCs, either their parent clouds or ambient interstellar material \citep{Dubner2015A&ARv,Kilpatrick2016ApJ,Su2017ApJ,Li2022MNRAS,Feng2024A&A}. CO line observations have proven highly effective in identifying such associations, with \citet{Zhou2023ApJS} finding that 80$\per$ of SNRs are associated with MCs. 
Importantly, the fundamental requirement of our method is a physical interaction between the SNR and an MC, making it inapplicable to remnants lacking such associations. Most Type \MakeUppercase{\romannumeral1}a SNRs explode in low-density environments lacking dense molecular gas \citep{Liu1997ApJ,Cooper2009ApJ}. Additionally, core-collapse supernovae resulting from high-velocity runaway stars may occur far from their parent clouds, making them unsuitable for this method \citep{Gies1987ApJS,Andersson2020MNRAS}.
Recent studies have successfully combined CO observations with 3D extinction maps to determine distances to both SNRs \citep{Chen2017MNRAS,Yu2019MNRAS,Li2022ApJ} and MCs \citep{Cao2023ApJ} by matching extinction features with molecular gas structures. In this work, we extend these approaches by using CO velocity information to first identify MCs physically interacting with SNRs, then employing 3D extinction maps to determine the distances to these associated clouds. By requiring both spectroscopic evidence of interaction and morphological correspondence between CO emission and extinction features, we can reliably adopt the MC distance as the SNR distance. This method provides robust distance estimates while minimizing contamination from unrelated clouds along the sightline.

The paper is organized as follows: Section \ref{sec:2data} describes the datasets, Section \ref{sec:method} outlines our methodology, Section \ref{sec: result} presents results of four SNRs and compares different distance estimation approaches, Section \ref{sec: discu} presents the discussion and Section \ref{sec: sum} summarizes our work.

\section{Data} \label{sec:2data}
We selected G93.7$-$0.2, G109.1$-$1.0, G156.2$+$5.7, and G166.0$+$4.3 from \citet{Zhao2020ApJ}, which have large angular sizes and are well-covered by both 3D extinction map and CO molecular line surveys. These SNRs span a range of morphological types and physical characteristics, forming a representative sample for testing our methodology. 

\subsection{CO Line Data}
The CO line emission in this work is from the Milky Way Imaging Scroll Painting (MWISP) CO survey project$\footnote{http://www.radioast.nsdc.cn/mwisp.php}$ \citep{Yang2026ApJS}, observed by the Purple Mountain Observatory Delingha 13.7 m millimeter telescope. The MWISP project covers a large area of 2310 deg$^{2}$ in the region of $l$ = $+9.75^{\circ}$ to $+229.75^{\circ}$ and $|b| \lesssim 5.25^{\circ}$. The CO data have a spatial resolution of approximately 50$^{\prime\prime}$, and the MWISP project contains the $\CO/\COO/\COOO$ ($J$ = 1$-$0) emission lines with a velocity resolution of $\sim$0.16 (0.17) $\kms$ for $\CO$ ($\COO$ and $\COOO$) lines. To investigate SNR-MC associations, the MWISP $\CO$ ($J$ = 1$-$0) data of G93.7$-$0.2, G109.1$-$1.0, and G166.0+4.3 are taken. Since G156.2+5.7 is beyond the coverage of the MWISP CO survey, its CO emission data are obtained from the CfA 1.2 m millimeter telescope $\CO$ ($J$ = 1$-$0) survey$\footnote{https://lweb.cfa.harvard.edu/rtdc/CO/}$ \citep{Dame2001ApJ}. The CfA CO survey provides a CO emission map with a coarse angular resolution of $8.5^{\prime}$ in latitude $|b| \lesssim 30^{\circ}$. 

\subsection{Three-Dimensional Extinction Map}
We employ the 3D dust extinction map of \citet{Green2019ApJ}, which covers three-quarters of the sky north of $\delta > -30^{\circ}$. This map combines stellar photometry from Gaia DR2, Pan-STARRS 1, and 2MASS to infer distances and extinctions to individual stars using a Bayesian framework. The map employs an adaptive multi-scale HEALPix tessellation scheme, achieving angular resolutions ranging from 3.4$^\prime$ to 13.7$^\prime$ depending on the local stellar density, with resolution decreasing at larger distances where stellar tracers become sparse. We access the extinction data through the Python package dustmaps\footnote{\url{https://dustmaps.readthedocs.io/}}, using the BayestarQuery function to extract extinction values along specific lines of sight. 

\subsection{Radio continuum observations} \label{radio}
The radio continuum images used in this study are obtained from the Canadian Galactic Plane Survey (CGPS) at 1420 MHz for G93.7$-$0.2, G109.1$-$1.0, and G166.0+4.3 \citep{Taylor2003AJ}, and from the Sino-German 6 cm polarization survey for G156.2+5.7 \citep{Sun2007A&A,Xu2007A&A}. These data are publicly available through the SNR catalog\footnote{\url{http://snrcat.physics.umanitoba.ca/SNRtable.php}} maintained by \citet{Ferrand2012AdSpR}. The angular sizes and basic properties of our sample are summarized in Table \ref{table1}. 

\section{Method} \label{sec:method}

We employ a two-step approach to determine SNR distances by finding out the MCs physically associated with SNRs and measuring their distances through 3D extinction mapping. This method relies on the fundamental physical constraint that an SNR and its interacting MCs must be at the same distance.

\begin{figure}[h!]
\centering
\includegraphics[width=0.8\textwidth]{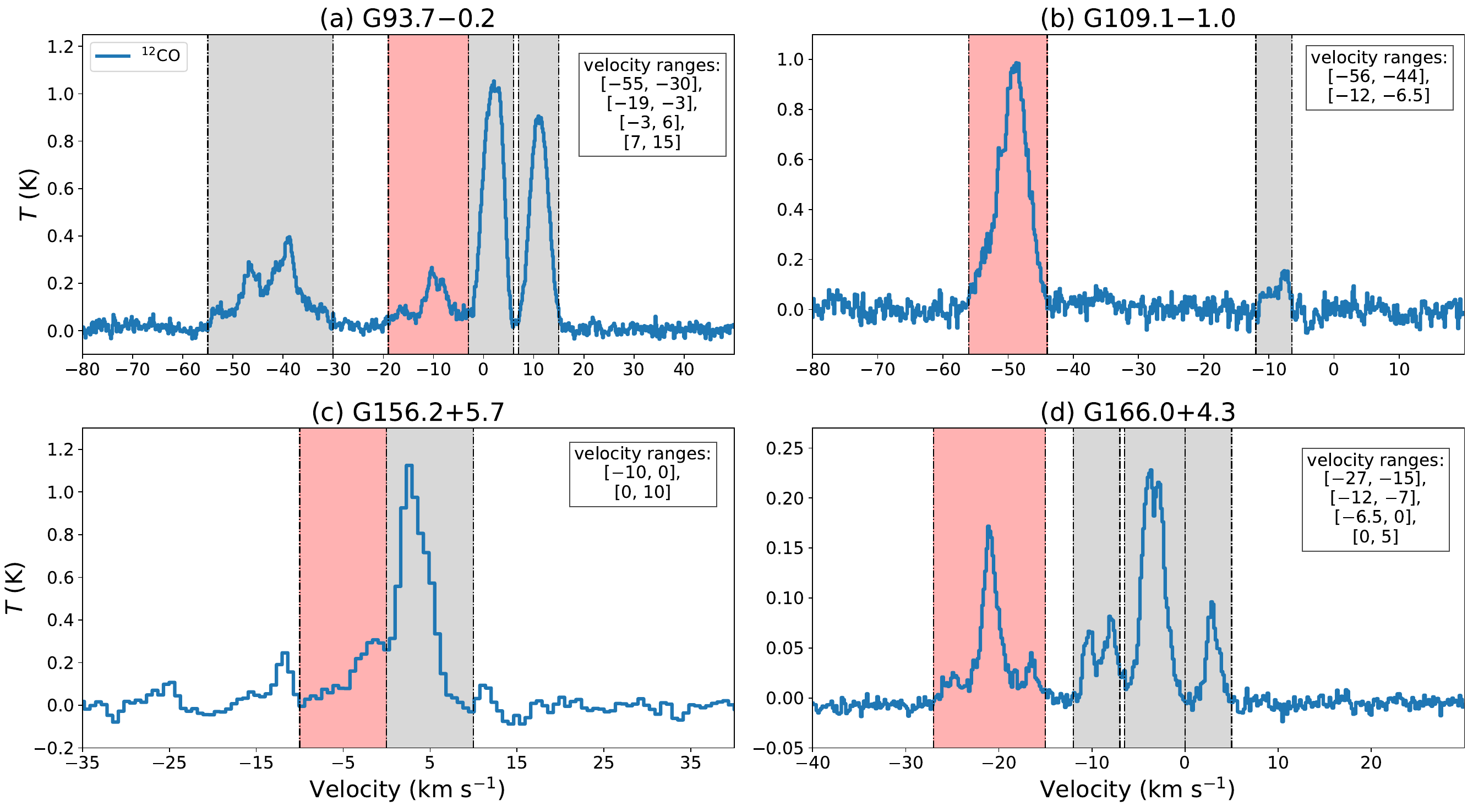}  
\caption{The averaged $\CO$ ($J$ = 1$-$0) spectra extracted from the regions of G93.7$-$0.2, G109.1$-$1.0, G156.2+5.7, and G166.0+4.3 by blue lines. Red shaded regions indicate the velocity components identified as associated with SNR-MC interactions, while gray shaded regions denote other unassociated velocity components along the sightline.}
\label{fig1}
\end{figure}

The first step is to identify SNR-associated MCs. We identify MCs associated with each SNR through analysis of $^{12}$CO emission data. For each SNR, we extract the average $^{12}$CO spectrum within the aperture defined in Table \ref{table1} and identify distinct velocity component corresponding to individual MC along the sightline. 
The velocity ranges of clouds interacting with each SNR are adopted from previous studies. These studies identified SNR-MC interactions primarily through robust kinematic evidence, such as the detection of broadened molecular lines (e.g., wings in CO line spectra) or shocked gas tracers. Morphological correspondence is considered in these studies only as supporting evidence.
Previous measurements of the SNR systemic velocities published in the literature are summarized (see references in Table \ref{table1}). Figure \ref{fig1} presents the extracted $^{12}$CO spectra for all four SNRs. Velocity ranges of interacting clouds are highlighted in red, while other components are shown in gray. For each velocity component, we generate velocity-integrated intensity maps over the corresponding ranges listed in Table \ref{table1}.

The next step is to determine the distance through 3D extinction map. For each velocity component, we select several positions with representative $^{12}$CO emission and extract their reddening $E(B-V)$ versus distance profiles. Sharp increases in these profiles indicate intervening dust structures. We identify distance ranges containing significant extinction jumps by requiring consistency across multiple sightlines. These ranges define distance bins for constructing differential extinction maps, where we calculate $\Delta E(B-V)$ for each spatial pixel between the bin boundaries. We then compare the morphology of these differential extinction maps with the velocity-integrated intensity maps. Based on the cloud boundaries, intensity peak positions, and overall morphology, the distance to the MC and associated SNR is determined by identifying the extinction map that exhibits the best spatial correspondence with CO emission.

\section{Result} \label{sec: result}
\subsection{G93.7$-$0.2 (CTB 104A, DA 551)} \label{G93.7}
G93.7$-$0.2 is a relatively large shell-type SNR with an angular diameter of approximately 80$\arcmin$. G93.7$-$0.2 has been extensively studied in radio continuum and H\MakeUppercase{\romannumeral1} observations \citep{Uyaniker2002ApJ}, which also exhibits strong polarization characteristics \citep{Uyaniker2003ApJ,Kothes2006A&A,Gao2011A&A} and displays a distorted thick-shell morphology in radio images. \citet{Uyaniker2002ApJ} identified a clear spatial correlation between G93.7$-$0.2 and neutral hydrogen gas at $-$6 \kms, suggesting a physical association between the remnant and surrounding H\MakeUppercase{\romannumeral1} structures.

\begin{figure}[ht!]
\centering
\includegraphics[width=0.8\textwidth]{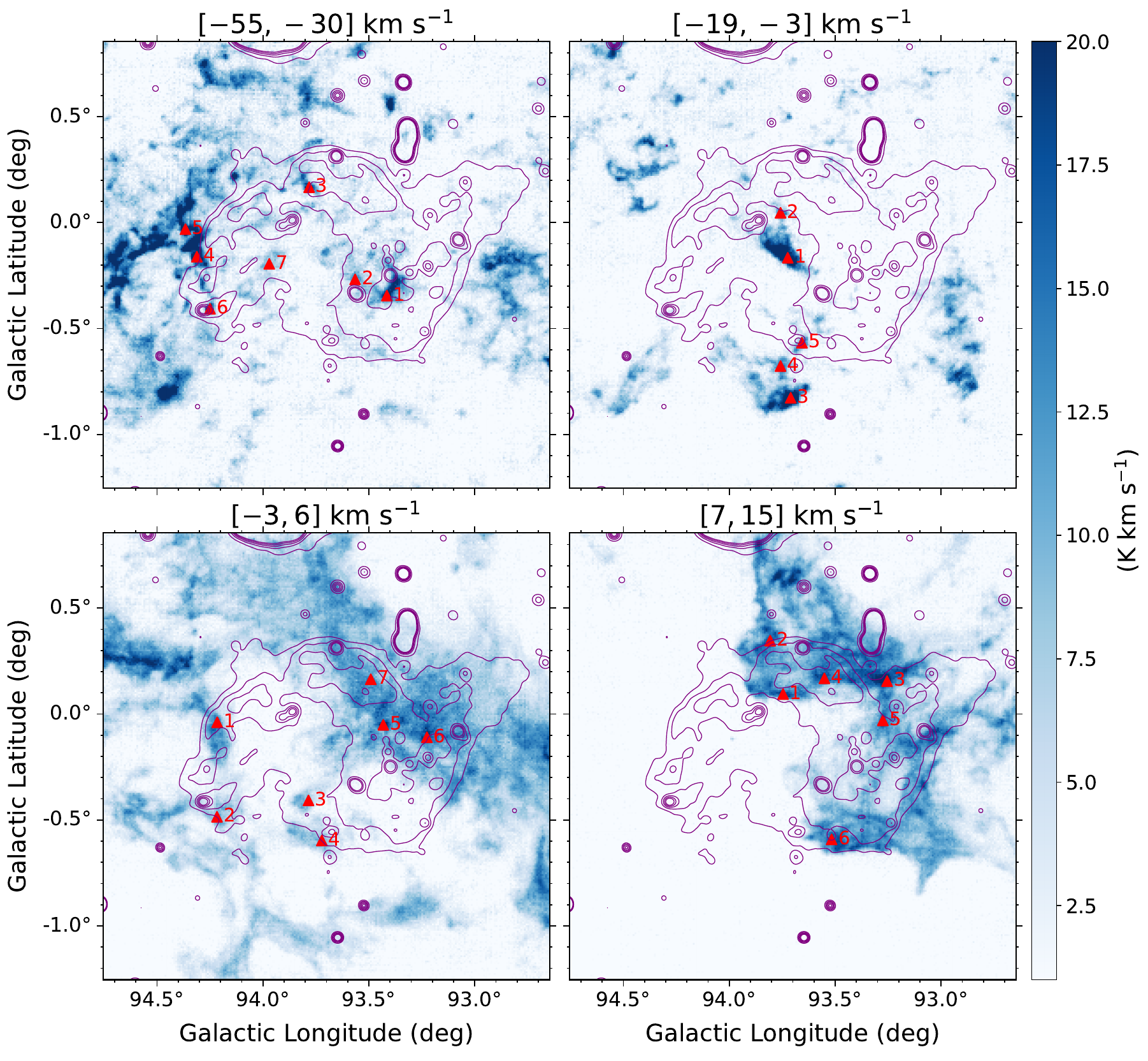}
\caption{Integrated intensity maps of $\CO$ for G93.7$-$0.2 corresponding to the velocity ranges of $[-55, -30]$ $\kms$, $[-19, -3]$ $\kms$, $[-3, 6]$ $\kms$, and [7, 15] $\kms$ respectively (see Figure \ref{fig1}(a)). Purple contours represent the 1420 MHz radio continuum emission, with the levels from 7.5 to 9.5 K with an interval of 0.5 K. Red triangles mark the representative $\CO$ emission positions selected for extracting extinction-distance profiles.}
\label{fig2}
\end{figure}

\begin{figure}[ht!]
\centering
\includegraphics[width=0.75\textwidth]{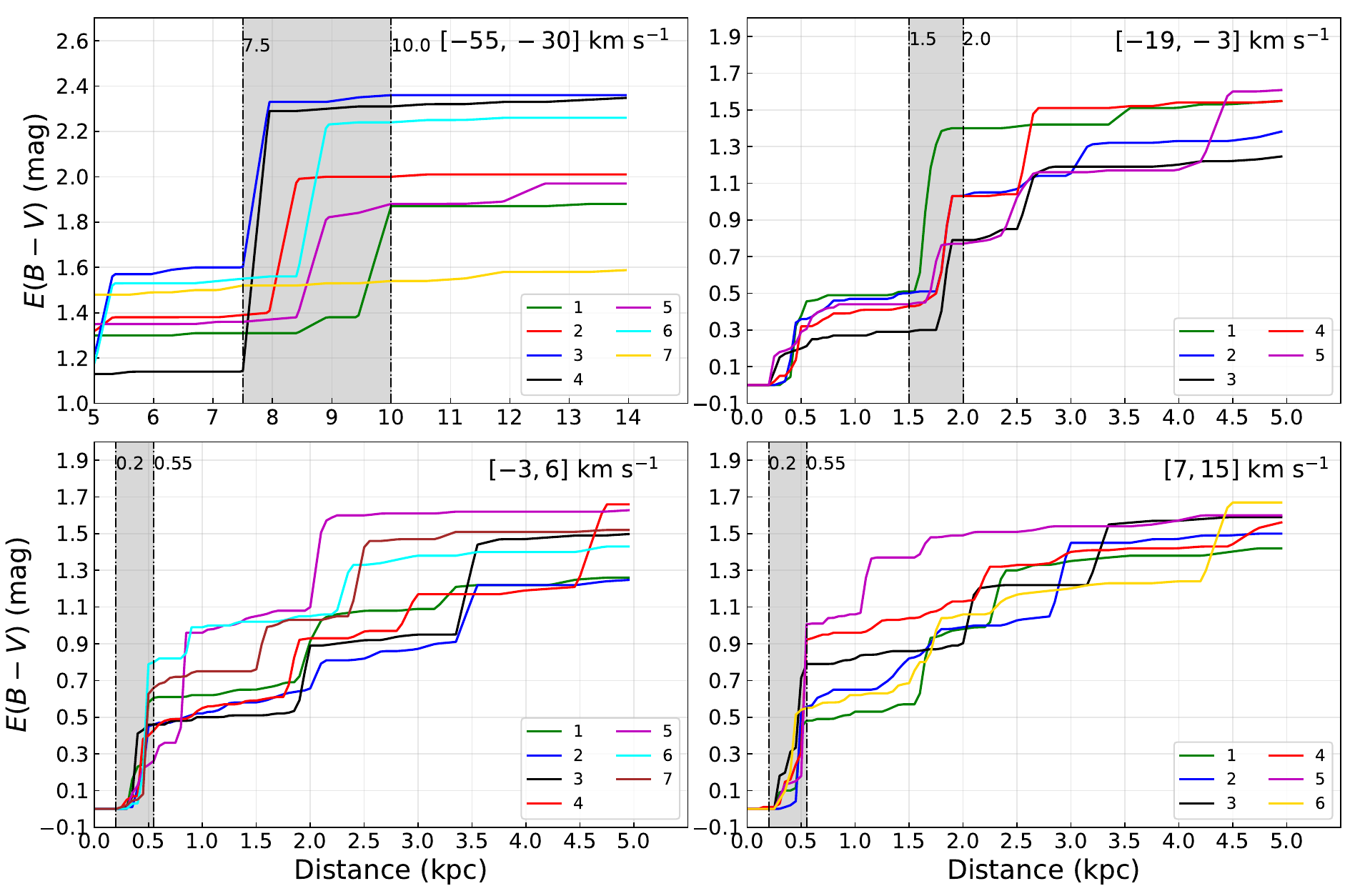}
\caption{The color excess $\EBV$ as a function of distance for G93.7$-$0.2 derived from 3D extinction map. Different colored lines correspond to the red positions marked in Figure \ref{fig2}. Vertical gray shaded regions indicate the extinction jump for the corresponding MC: 7.5--10.0 kpc for $[-55, -30]$ $\kms$, 1.5--2.0 kpc for $[-19, -3]$ $\kms$, and 0.2--0.55 kpc for both $[-3, 6]$ $\kms$ and $[7, 15]$ $\kms$.}
\label{fig3}
\end{figure}

From the MWISP $^{12}$CO data, we revealed four distinct velocity components in the G93.7$-$0.2 region: [$-$55, $-$30] \kms, [$-$19, $-$3] \kms, [$-$3, 6] \kms, and [7, 15] \kms, as shown in Figure \ref{fig1}(a). The velocity-integrated intensity maps of these components are presented in Figure \ref{fig2}. These MCs overlap along the sightline and exhibit diffuse morphologies. \citet{Zhou2023ApJS} analyzed the same CO data and found no clear evidence for SNR-MC interaction. Following the argument of \citet{Uyaniker2002ApJ}, we tentatively identify the [$-$19, $-$3] \kms\ component as potentially interacting with G93.7$-$0.2. To determine distances to these MCs, we select positions with representative CO emission from each velocity component and extract extinction-distance profiles derived from the 3D extinction map, shown in Figure \ref{fig3}. All four velocity components exhibit a prominent extinction jump at $\sim$ 0.5 kpc, likely associated with the edge of the Local Bubble \citep{Wang2025ApJS}. The [$-$55, $-$30] \kms\ component shows an additional significant jump at $\sim$ 7.5 kpc, while the [$-$19, $-$3] \kms\ component displays a concentrated jump at $\sim$ 1.8 kpc. In contrast, the [$-$3, 6] \kms\ and [7, 15] \kms\ components show no additional significant jumps beyond 0.5 kpc. The gray shades in Figure \ref{fig3} indicate extinction jumps associated with each velocity component.

\begin{figure}[ht!]
\centering
\includegraphics[width=0.75\textwidth]{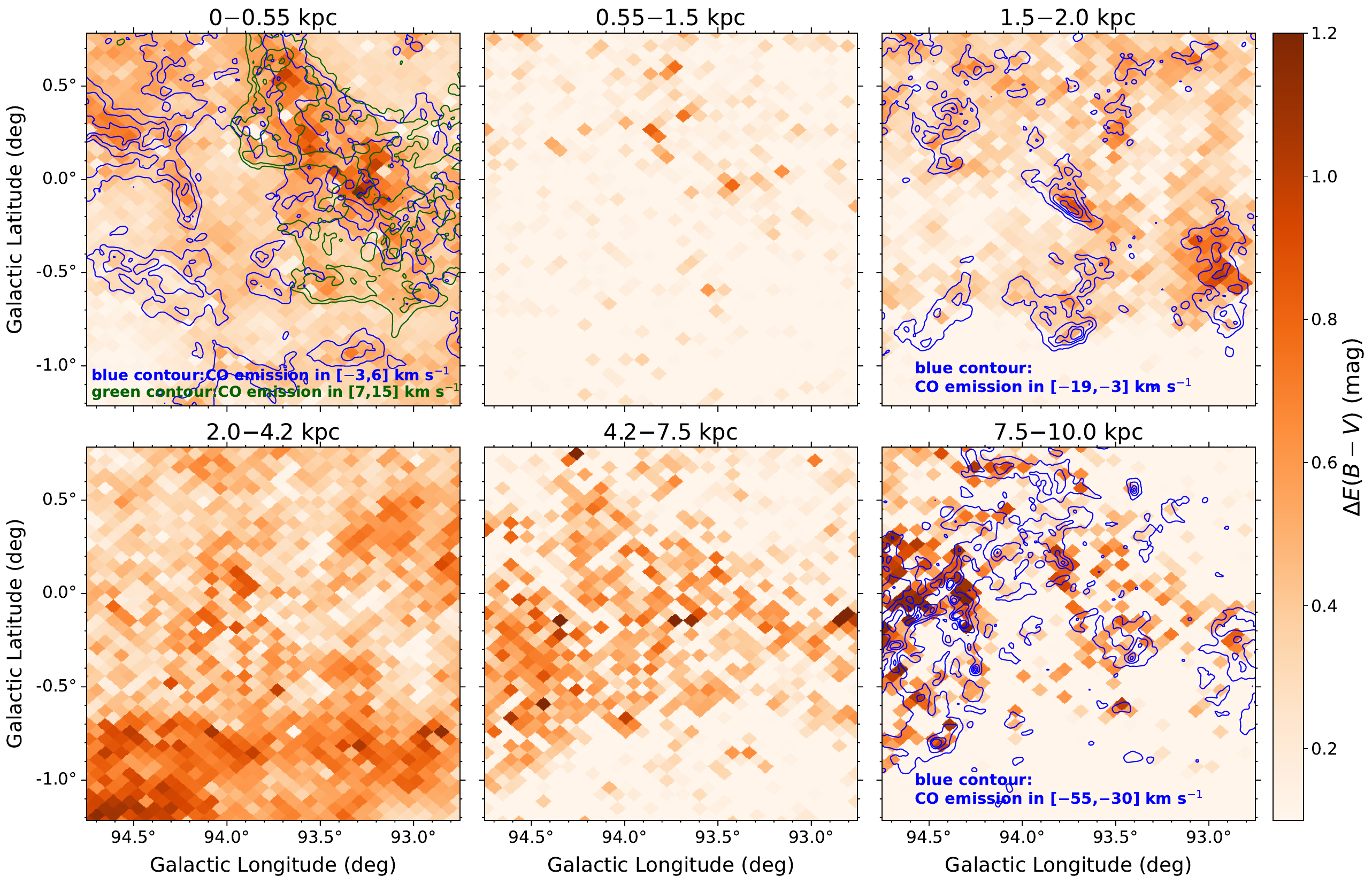}
\caption{Differential extinction maps ($\Delta E(B-V)$) toward G93.7$-$0.2 for the distance bins of 0--0.55 kpc, 0.55--1.5 kpc, 1.5--2.0 kpc, 2.0--4.2 kpc, 4.2--7.5 kpc, and 7.5--10.0 kpc respectively. Blue contours in the 1.5--2.0 kpc and 7.5--10.0 kpc bins represent $\CO$ emission integrated over $[-19, -3]$ $\kms$ and $[-55, -30]$ $\kms$ respectively. In the 0--0.55 kpc bin, contours show $\CO$ emission for the [7, 15] $\kms$ (green) and $[-3, 6]$ $\kms$ (blue) components. Contours are plotted at levels of 6.0, 13.0, 21.0, and 28.0 $\kkms$ for $[-55, -30]$ $\kms$ component; 3.0, 10.0, 17.0, and 24.0 $\kkms$ for $[-19, -3]$ $\kms$ component; 4.5, 9.5, and 14.5 $\kkms$ for $[-3, 6]$ $\kms$ component; 4.5, 11.5, and 18.5 $\kkms$ for [7, 15] $\kms$ component. The color bar indicates $\Delta E(B-V)$ values ranging from 0.1 to 1.2 mag.}
\label{fig4}
\end{figure}

These extinction jump intervals provide the basis for dividing the 3D extinction map. In Figure \ref{fig3}, the differential extinction from 0 to 0.2 kpc is nearly 0 mag, which is attributable to the Local Bubble. Therefore, this range does not affect the differential extinction maps. Consequently, we define the first extinction map from 0 to 0.55 kpc.
Within the broad distance interval from 2.0 to 7.5 kpc, the extinction profile exhibits dispersed and low-amplitude jumps, and no associated CO emission components are identified within this distance interval. Given the inherent distance uncertainties in the extinction map and the complex interstellar environment of SNR, we adopt 4.2 kpc as a dividing boundary and assign all dispersed extinction jumps occurring prior to this distance to a single extinction bin.
Finally, we divide the 3D extinction map into the following distance bins: 0--0.55 kpc, 0.55--1.5 kpc, 1.5--2.0 kpc, 2.0--4.2 kpc, 4.2--7.5 kpc, and 7.5--10.0 kpc. Figure \ref{fig4} compares the spatial distributions of CO emission for each velocity component with the differential extinction maps in these distance bins. The [$-$55, $-$30] \kms\ component shows the high spatial correlation with the extinction map in the 7.5--10.0 kpc bin, though we note that the 3D extinction map may have significant uncertainties at such large distances. The [$-$19, $-$3] \kms\ component exhibits excellent morphological agreement with the extinction map in the 1.5--2.0 kpc bin, suggesting a distance of $\sim$1.8 kpc. Both the [$-$3, 6] \kms\ and [7, 15] \kms\ components correlate well with the 0--0.55 kpc extinction map, indicating they likely represent a single molecular complex at $\sim$ 0.5 kpc.

If the [$-$19, $-$3] \kms\ component is indeed interacting with G93.7$-$0.2, we can constrain the SNR distance to approximately 1.8 kpc. This value is slightly smaller than recent extinction-based distances of 1.99$\pm$0.33 kpc \citep{Wang2020AA} and 2.16$\pm$0.02 kpc \citep{Zhao2020ApJ}, but larger than the kinematic distance of 1.5$\pm$0.2 kpc derived from H\MakeUppercase{\romannumeral1} data \citep{Uyaniker2002ApJ}. \citet{Ranasinghe2022ApJ} recently revised the kinematic distance to 1.3$\pm$0.2 kpc using updated rotation curves. Our methodology provides a more robust distance determination of $1.82 \pm 0.13$ kpc, which corresponds to a physical diameter of about 35 pc for G93.7$-$0.2. All results are summarized in Table \ref{table2}.
We determine the distance to G93.7$-$0.2 from the differential extinction per distance $\Delta E(B-V)/D$ with distance $D$. Specifically, the location of a sharp gradient that is significantly larger than the average is adopted as the SNR distance. The boundaries of the 1$\sigma$ confidence interval were taken as the 16$\per$ and 84$\per$ of gradient peak, which are taken as distance uncertainties. The same approach is applied consistently to all subsequent SNRs in our sample.

\subsection{G109.1$-$1.0 (CTB 109)} \label{G109.1-1.0} 
G109.1$-$1.0 exhibits a semicircular shell morphology in both radio and X-ray bands, spanning approximately 30$^\prime$ \citep{Gregory1980Nature,Hughes1981}. This morphology likely results from interaction with adjacent MCs \citep{Tatematsu1987A&A}. The anticorrelation between the X-ray emission and CO cloud morphology provides evidence for SNR-MC interaction \citep{Sasaki2004ApJ,Kothes2012}. The remnant hosts the anomalous X-ray pulsar AXP J2301+5852 (1E 2259+586) near its geometric center \citep{Olausen2014}. Previous studies have identified molecular gas at $-$52 to $-$50 \kms\ associated with G109.1$-$1.0, placing it in the Perseus spiral arm \citep{Kothes2002ApJ}. However, some observations found no clear evidence of shocked CO \citep{Tatematsu1990ApJ,Zhou2023ApJS}, and subsequent studies detected weak line broadening near [$-$57, $-$52] \kms\ \citep{Sasaki2006ApJ}. Recent HCO$^+$ observations reveal a blue-shifted line wing around $-$50 \kms, potentially indicating shock interaction \citep{Tu2025}. 

\begin{figure}[ht!]
\centering
\includegraphics[width=0.8\textwidth]{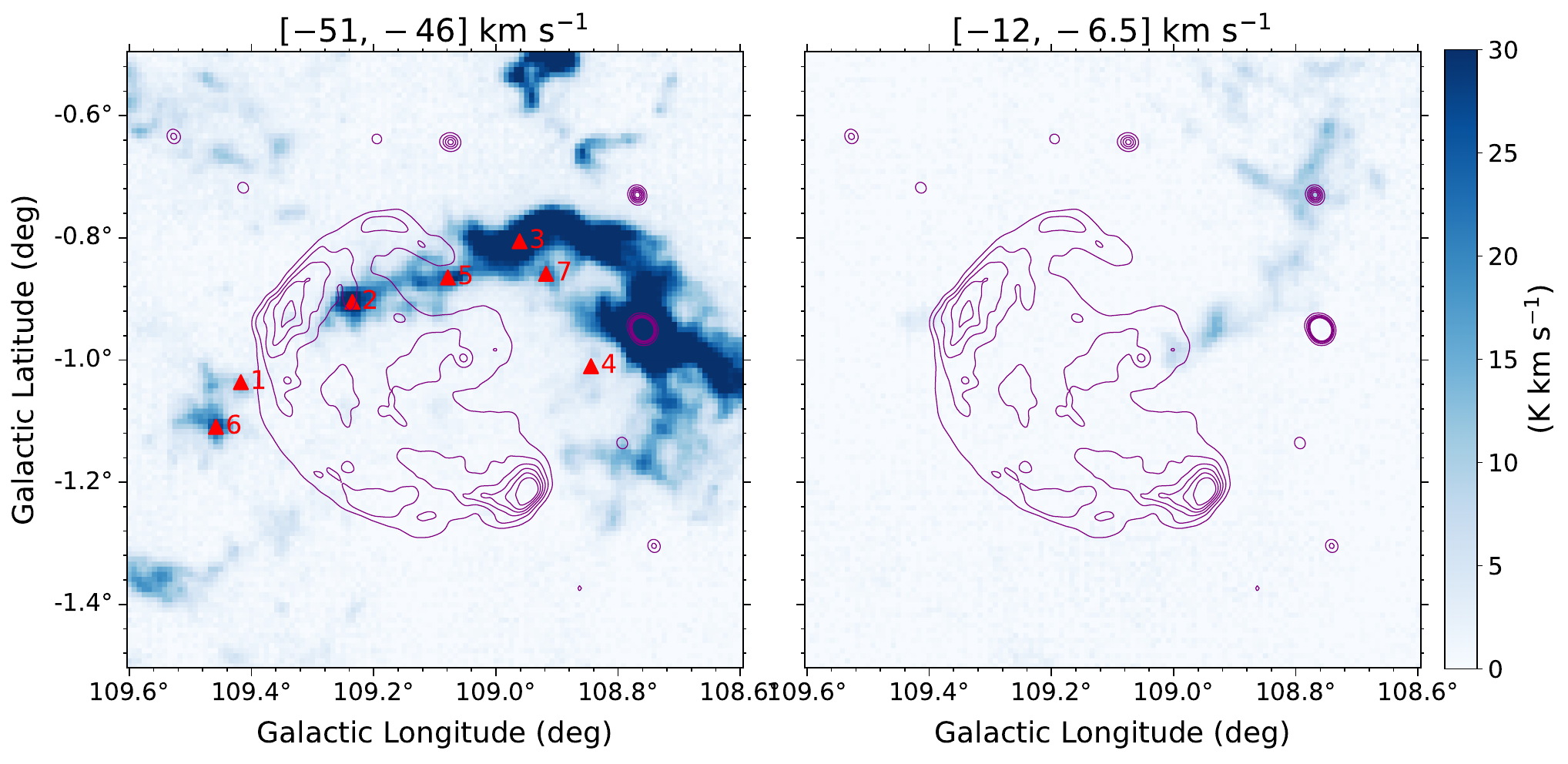}
\caption{Integrated intensity maps of $\CO$ for G109.1$-$1.0 corresponding to the $[-51, -46]$ $\kms$ and $[-12, -6.5]$ $\kms$ components (see Figure \ref{fig1}(b)). Purple contours represent the 1420 MHz radio emission shell of SNR, with the levels from 0 to 24.0 K with an interval of 3.0 K. The red triangles are used for the same purpose as in Figure \ref{fig2}.}
\label{fig5}
\end{figure}

Our MWISP $\CO$ observations reveal two distinct velocity components toward G109.1$-$1.0: [$-$56, $-$44] \kms\ and [$-$12, $-$6.5] \kms, with the former showing significantly stronger emission in Figure \ref{fig1}(b). However, the [$-$56, $-$44] \kms\ component exhibits a weak peak near $-$51 \kms\, indicating that this velocity range likely consists of multiple molecular structures. In addition, the massive MC to the west of SNR blocks the shock wave from further expansion, and the MC's extended bridge-shape structure (CO ridge) is likely in front of the SNR or interacting with it \citep{Tatematsu1990ApJ,Kothes2002ApJ}. The CO velocity range in which both the MC and CO ridge are present is therefore crucial for identifying SNR-MC interaction. Based on this evidence for interaction, we adopt the [$-$51, $-$46] \kms\ component as physically associated with the SNR. Figure \ref{fig5} presents CO integrated intensity maps for both components overlaid with radio continuum contours. The [$-$51, $-$46] \kms\ component exhibits clear morphological anticorrelation with the radio shell.

\begin{figure}[ht!]
\centering
\includegraphics[width=0.8\textwidth]{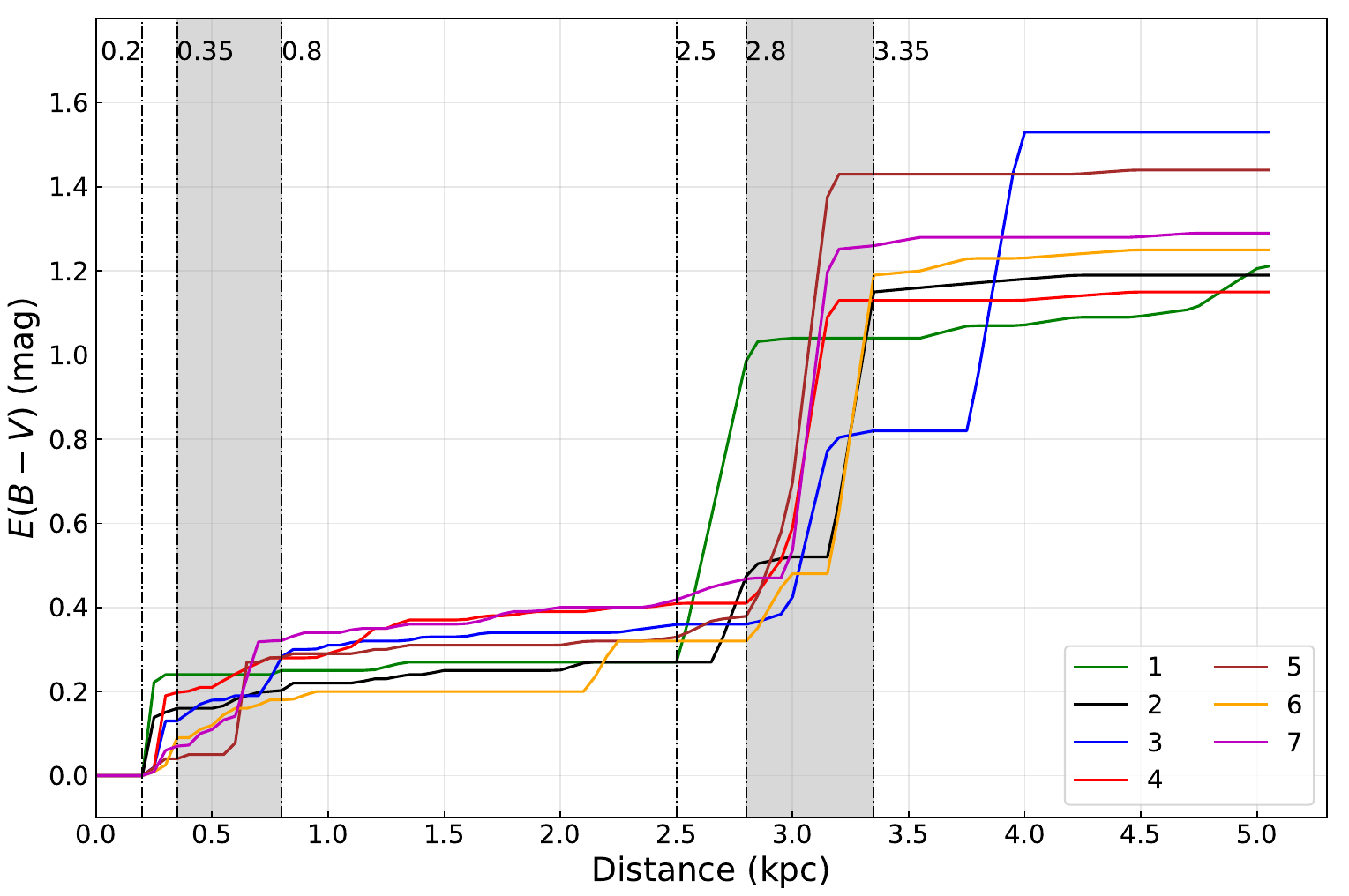}
\caption{The color excess $\EBV$ as a function of distance for G109.1$-$1.0 derived from 3D extinction map. Different colored lines correspond to the red positions marked in Figure \ref{fig5}. The right gray shaded region (2.8--3.35 kpc) indicates the extinction jump caused by the SNR-associated MC, while the left gray shaded region (0.35--0.8 kpc) corresponds to a foreground cloud.}
\label{fig6}
\end{figure}

\begin{figure}[ht!]
\centering
\includegraphics[width=0.8\textwidth]{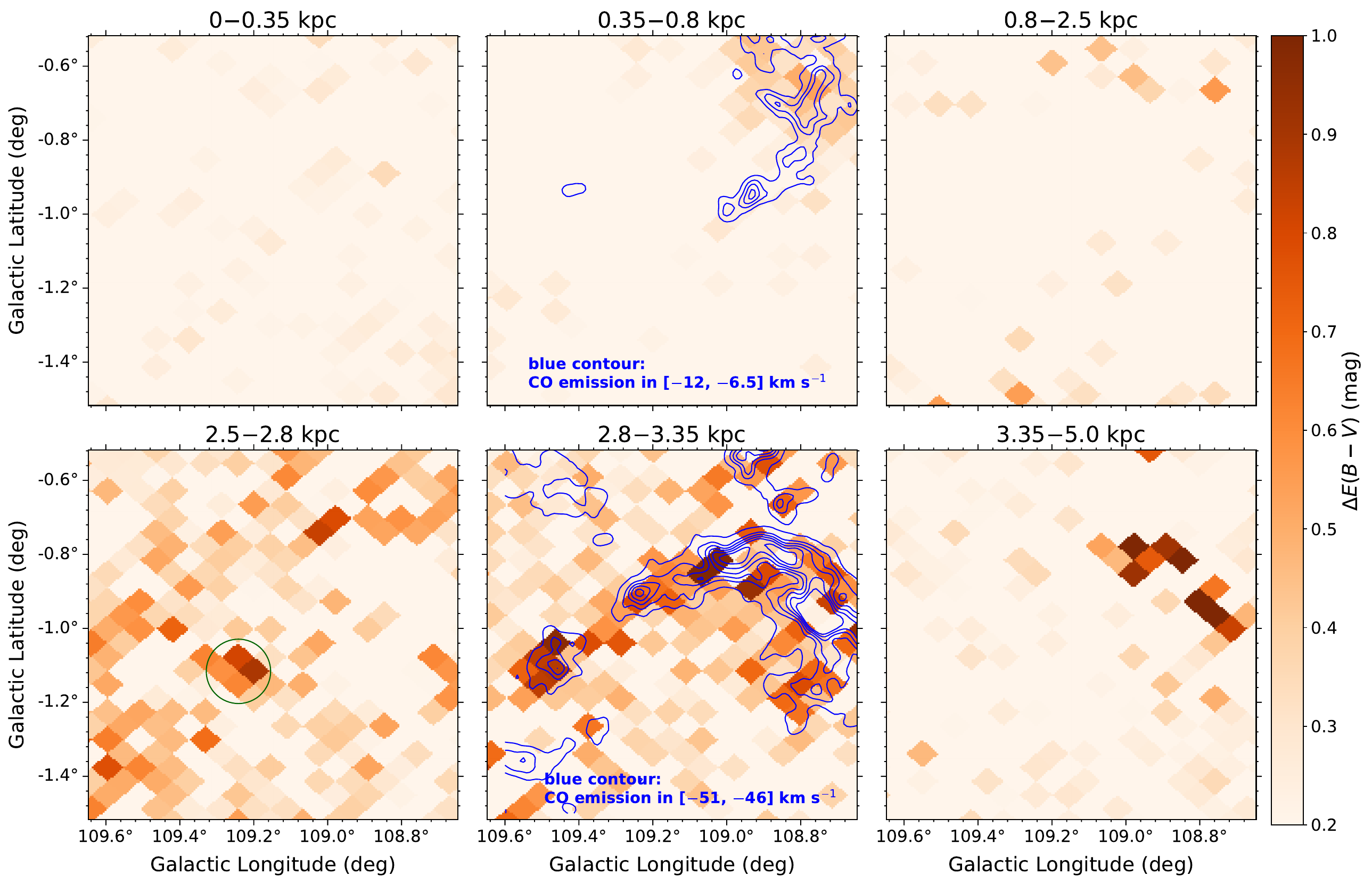}
\caption{Differential extinction maps ($\Delta E(B-V)$) toward G109.1$-$1.0 for distance bins of 0--0.35 $\kpc$, 0.35--0.8 $\kpc$, 0.8--2.5 $\kpc$, 2.5--2.8 $\kpc$, 2.8--3.35 $\kpc$, and 3.35--5.0 $\kpc$ respectively. Blue contours overlay $\CO$ emission for the $[-12, -6.5]$ $\kms$ component (0.35--0.8 kpc bin) and the $[-51, -46]$ $\kms$ component (2.8--3.35 kpc bin) respectively. Contours are plotted at levels of 3.0, 10.0, 17.0, 24.0, 31.0, and 38.0 $\kkms$ for $[-51, -46]$ $\kms$ component; 2.0, 5.0, 8.0, and 11.0 $\kkms$ for $[-12, -6.5]$ $\kms$ component. The green circle in the 2.5--2.8 kpc bin marks a molecular clump located adjacent to and in front of the massive MC. The color bar indicates $\Delta E(B-V)$ values ranging from 0.2 to 1.0 mag.}
\label{fig7}
\end{figure}

We extracted extinction-distance profiles at multiple positions with representative CO emission in the [$-$51, $-$46] \kms\ component, as shown in Figure \ref{fig6}. These profiles show prominent extinction jumps at approximately 0.7 kpc and 3.0 kpc. 
We divided the extinction data into six distance bins: 0--0.35 kpc, 0.35--0.8 kpc, 0.8--2.5 kpc, 2.5--2.8 kpc, 2.8--3.35 kpc, and 3.35--5.0 kpc, generating differential extinction maps for each bin in Figure \ref{fig7}. The 0.35--0.8 kpc extinction map correlates well with the [$-$12, $-$6.5] \kms\ component, indicating a foreground cloud at $\sim$0.7 kpc. A molecular clump is identified within the green aperture at 2.5--2.8 kpc bin. This clump is not part of the massive MC. The extinction map of 2.8--3.35 kpc bin shows excellent morphological agreement with the [$-$51, $-$46] \kms\ component, constraining G109.1$-$1.0 to a distance of approximately 3.0 kpc. 

This distance is consistent with most previous measurements, as summarized in Table \ref{table3}. \cite{Kothes2002ApJ} placed G109.1$-$1.0 at 3.0 $\pm$ 0.5 kpc in the Perseus arm based on kinematic arguments, including the systemic velocity $-$50 \kms\ of associated MC and comparison with nearby H\MakeUppercase{\romannumeral2} regions of known spectroscopic distances. \cite{Kothes2012} synthesized all available observational evidence and firmly established a distance for G109.1$-$1.0 of 3.2 $\pm$ 0.2 kpc, locating it within or near the spiral shock zone of Perseus arm. \cite{Sanchez2018MNRAS} derived 3.1 $\pm$ 0.2 kpc optical kinematics from H$\alpha$ and [S \MakeUppercase{\romannumeral2}] observations. Recent extinction-based studies by \cite{Zhao2020ApJ} and \cite{Chen2025ApJ} obtained 2.8 $\pm$ 0.04 kpc and $\sim$3.1 kpc respectively. A large distance of 7.5 $\pm$ 1.0 kpc estimated by \cite{Durant2006ApJ} has been criticized for methodological issues which do not account for the large intrinsic scatter \citep{Kothes2012}. \cite{Tian2010MNRAS} placed G109.1$-$1.0 at the far kinematic distance of 4 kpc based on the apparent absence of H\MakeUppercase{\romannumeral1} self-absorption (HISA) features associated with the MC complex. But their result is challenged by the detection of weak, patchy HISA features in the CGPS data. Given the strong evidence for SNR-MC interaction and the morphological correspondence between extinction and CO emission, we recommend a distance of $3.05 \pm 0.15$ kpc for G109.1$-$1.0.

\subsection{G156.2+5.7} \label{G156.2+5.7}
To constrain its distance and physical associations, previous studies have investigated the environment of giant (size: 110$^\prime$) SNR G156.2+5.7. \citet{Reich1992AA} identified an H\MakeUppercase{\romannumeral1} shell surrounding the remnant at velocities between $-48$ and $-41$ $\kms$ and noted a CO emission component spanning $-$8 to 8 $\kms$ in the SNR region. However, despite these results, a definitive identification of the specific MCs physically interacting with the SNR has remained uncertain.
 
\begin{figure}[ht!]
\centering
\includegraphics[width=0.75\textwidth]{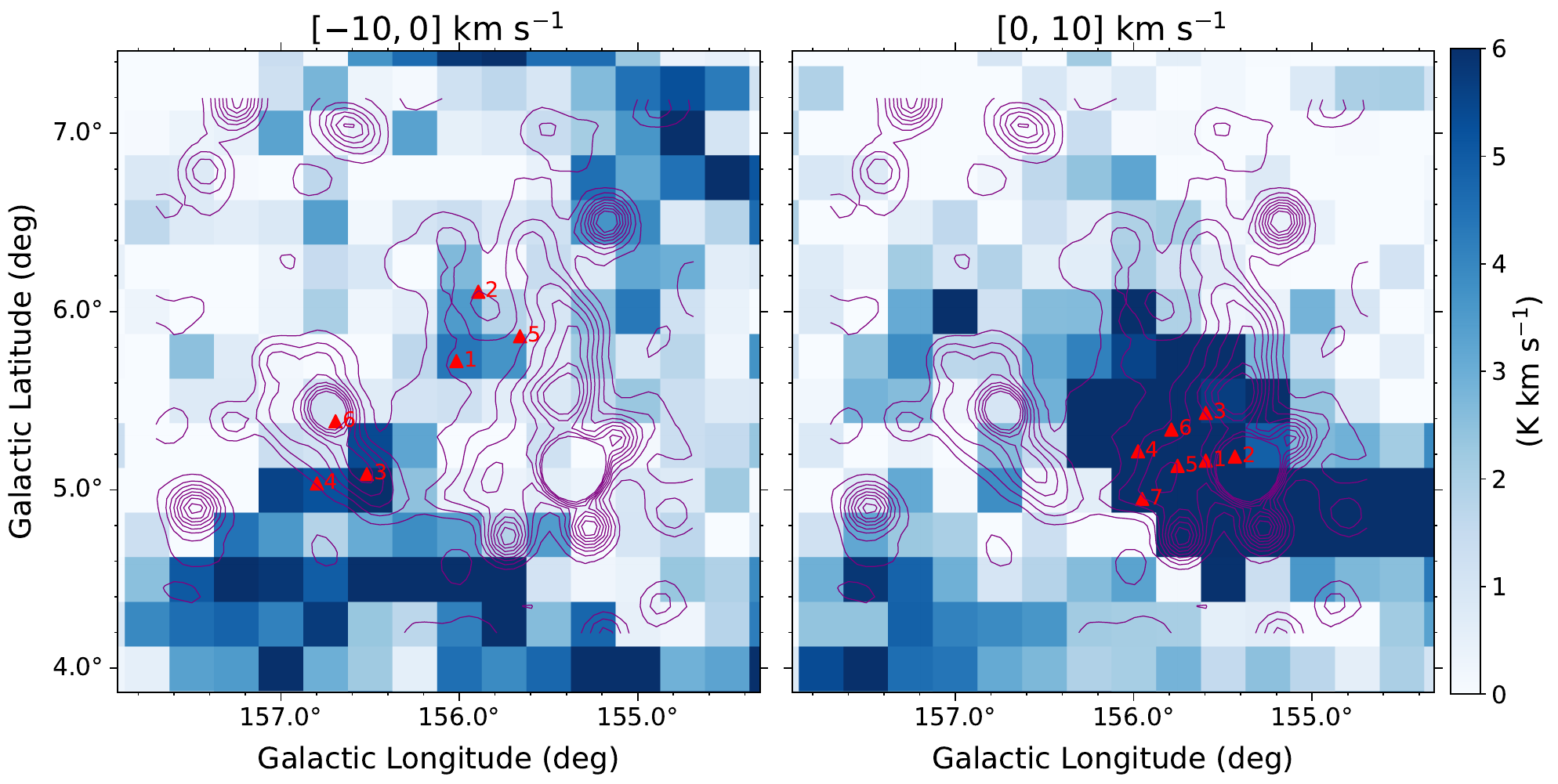}
\caption{Integrated intensity maps of $\CO$ for G156.2+5.7 over the $[-10, 0]$ $\kms$ and [0, 10] $\kms$ velocity components respectively (see Figure \ref{fig1}(c)). Purple contours represent the Sino-German 6 cm radio emission shell of SNR, with the levels from 1.0 to 15.0 mK with an interval of 2.0 mK. The red triangles are used for the same purpose as in Figure \ref{fig2}.}
\label{fig8}
\end{figure}

Using CfA $^{12}$CO emission data, we have revealed the molecular environment and identified two distinct velocity components toward G156.2+5.7: $[-10, 0]$ $\kms$ and [0, 10] $\kms$ in Figure \ref{fig1}(c). While the [0, 10] $\kms$ component partially overlaps the western part of the radio image in Figure \ref{fig8}, it lacks a clear morphological correspondence with the SNR shell. Consequently, we suggest that the positive velocity emission previously noted in the broader range by \citet{Reich1992AA} is likely a chance superposition unrelated to the SNR. In contrast, the [$-$10, 0] $\kms$ component exhibits a morphological agreement with the SNR, with its strong emission regions distributed along the edge of radio shell. This spatial correlation implies that the [$-$10, 0] $\kms$ component represents the site of SNR-MC interaction. 

\begin{figure}[ht!]
\centering
\includegraphics[width=0.8\textwidth]{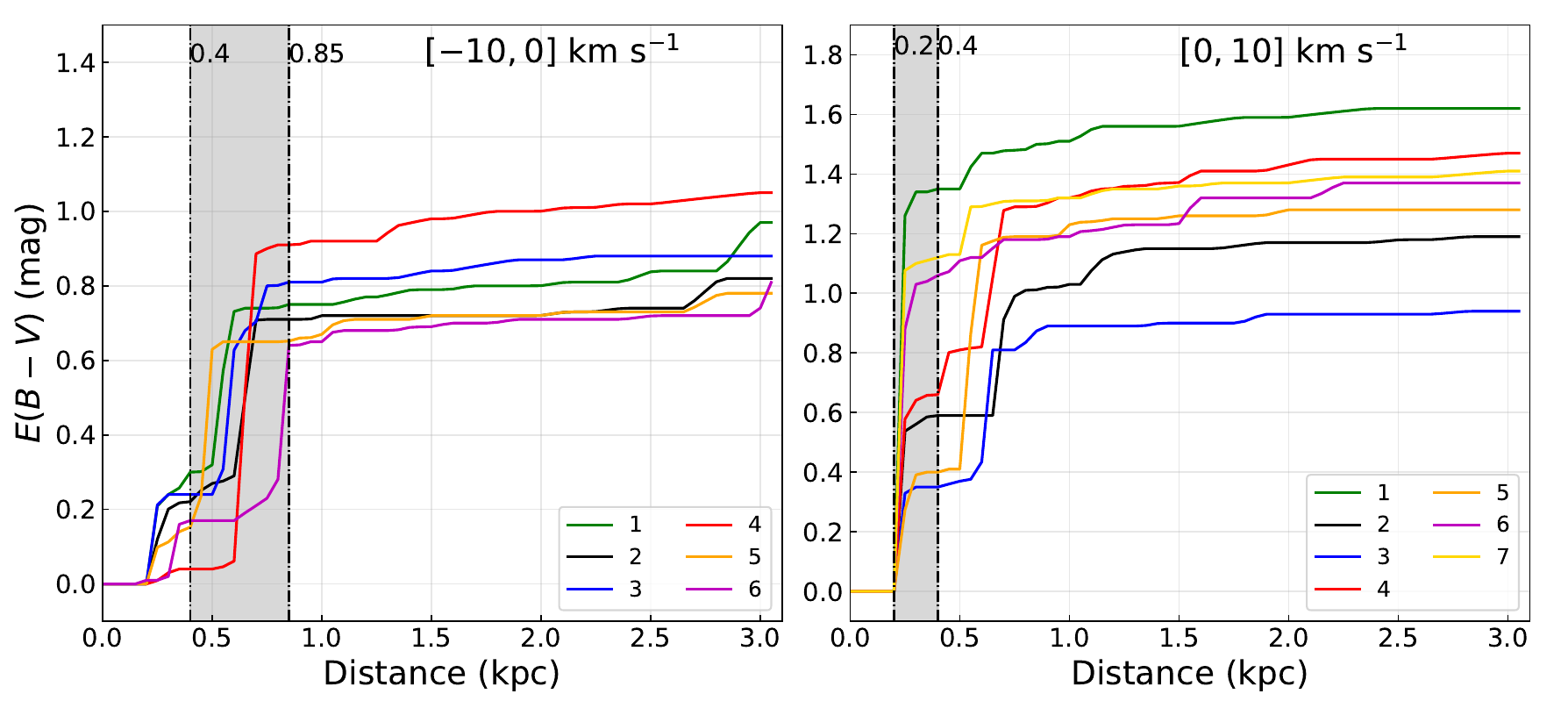}
\caption{The color excess $\EBV$ as a function of distance for G156.2+5.7 derived from 3D extinction map. Different colored lines correspond to the red positions marked in Figure \ref{fig8}. Vertical gray regions indicate extinction jumps for the $[-10, 0]$ $\kms$ component (0.4--0.85 kpc) and the [0, 10] $\kms$ component (0.2--0.4 kpc).}
\label{fig9}
\end{figure}

\begin{figure}[ht!]
\centering
\includegraphics[width=0.7\textwidth]{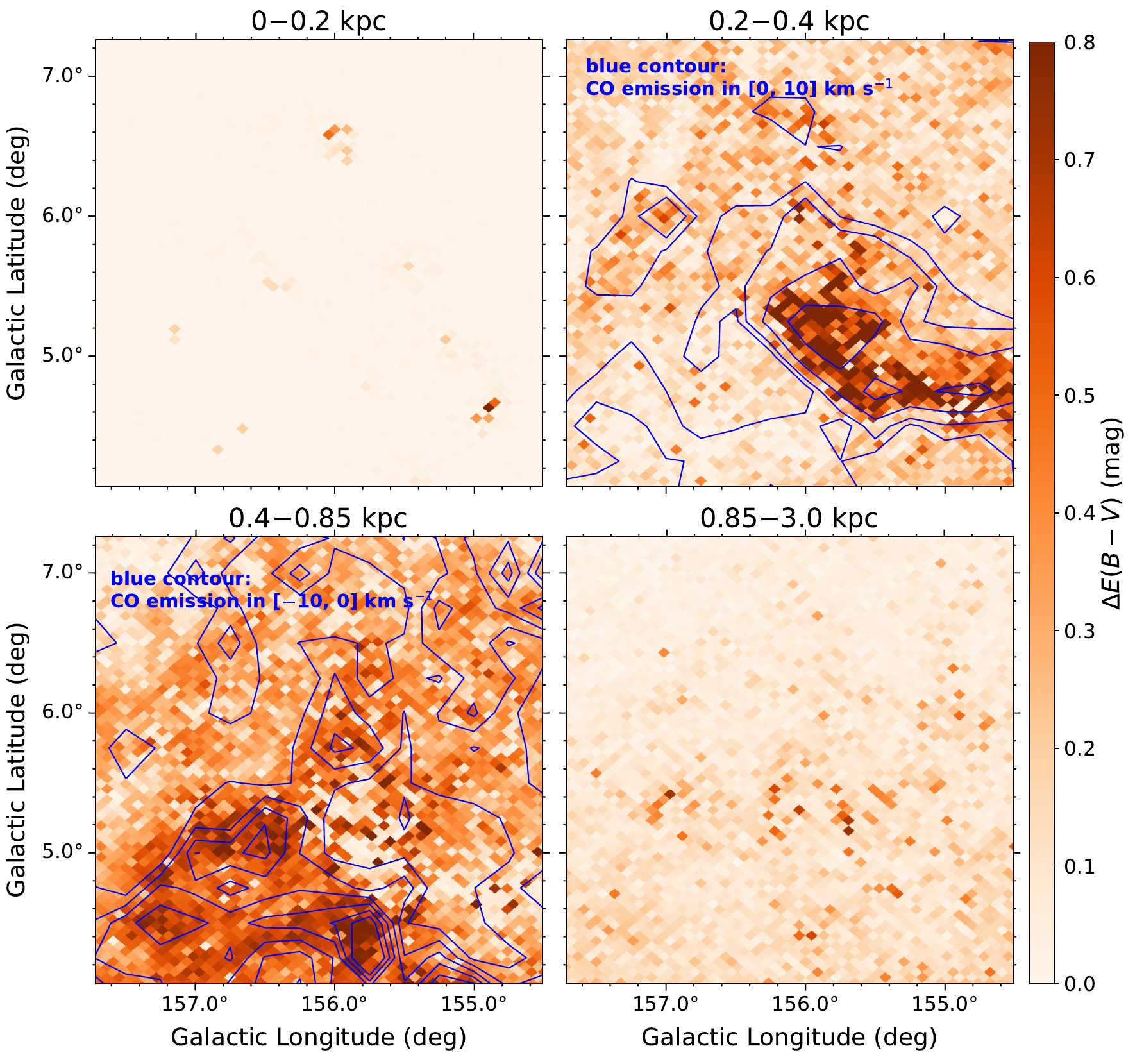}
\caption{Differential extinction maps ($\Delta E(B-V)$) toward G156.2+5.7 for distance bins of 0--0.2 kpc, 0.2--0.4 kpc, 0.4--0.85 kpc, and 0.85--3.0 kpc. Blue contours represent $\CO$ emission for the [0, 10] $\kms$ component (0.2--0.4 kpc bin) and the $[-10, 0]$ $\kms$ component (0.4--0.85 kpc bin) respectively. Contours are plotted at levels of 1.0, 2.5, 4.0, 5.5, 7.0, and 8.5 $\kkms$ for $[-10, 0]$ $\kms$ component; 2.0, 4.0, 8.0, and 14.0 $\kkms$ for $[0, 10]$ $\kms$ component. The color bar indicates $\Delta E(B-V)$ values ranging from 0 to 0.8 mag.}
\label{fig10}
\end{figure}

To determine the distance to this interacting cloud, we constructed extinction-distance profiles by the positions of representative $^{12}$CO emission in Figure \ref{fig8}. According to the extinction jumps in Figure \ref{fig9}, the MCs along the sightline toward G156.2+5.7 are primarily concentrated within 3.0 kpc, while the region beyond this distance exhibits negligible extinction. Consequently, our extinction map for this SNR focuses on the 0--3.0 kpc range. Given these extinction jumps occur at relatively nearby positions, we also retain the 0--0.2 kpc interval (corresponding to the Local Bubble) to provide detailed extinction maps of the MC along the sightline. Therefore, we divided the extinction maps into four distance bins: 0--0.2 kpc, 0.2--0.4 kpc, 0.4--0.85 kpc, and 0.85--3.0 kpc, as shown in Figure \ref{fig10}. The extinction jump observed at 0.2--0.4 kpc is likely caused by the dense boundary of the Local Bubble, a foreground feature unrelated to the SNR. However, the extinction structure in the 0.4--0.85 kpc bin shows a high morphological coincidence with the [$-$10, 0] $\kms$ velocity component. We therefore identify the 0.4--0.85 kpc bin as the location of the SNR-MC interaction, constraining the distance of G156.2+5.7 to approximately 0.6 kpc.

Our distance estimate of $\sim$0.6 kpc is consistent with the optical analysis of \citet{Gerardy2007MNRAS}, who inferred a distance of 0.3--0.6 kpc, as well as the extinction-based measurement of \citet{Zhao2020ApJ} who derived $0.68 \pm 0.2$ kpc, see in Table \ref{table4}. These results, which place the SNR within 1 kpc, contrast sharply with the substantially larger estimates like the kinematic distance (1--3 kpc) proposed by \citet{Reich1992AA}, the distance $\textgreater$1.7 kpc derived from proper motion and shock velocity \citep{Katsuda2016ApJ}, and X-ray based results (3 kpc from \citet{Pfeffermann1991A&A} and 1.3 kpc from \citet{Yamauchi1999PASJ}). Discrepancies among the X-ray distances likely originate from different assumptions regarding the local ambient density in Sedov models. Furthermore, \citet{Xu2007A&A} provided strong independent evidence that the SNR shock is interacting with surrounding dense interstellar clouds by radio and IRAS 100 $\mum$ emission correlations, and inferred a distance of 1 kpc by assessing the spatial location of SNR relative to the Galactic spiral arms. This 1 kpc estimate aligns closely with our proposed near distance range and supports the reliability of our overall interaction scenario. Our analysis refines this value to approximately 0.6 kpc because the 3D extinction map provides a more robust and accurate method for locating specific MCs along the sightline compared to large-scale spiral arm projections. Based on the complementary morphologies between the radio shell and the [$-$10, 0] $\kms$ component, we conclude that G156.2+5.7 is located at $0.60 \pm 0.15$ kpc.

\subsection{G166.0+4.3 (VRO 42.05.01)} \label{G166.0+4.3}
SNR G166.0+4.3 presents a peculiar morphology in the radio band, characterized by a smaller semi-circular ``shell" and a larger triangular ``wing", exhibiting an edge-brightened, double-shell structure \citep{Pineault1985A&A,Guo1997ApJ,Leahy2005A&A}. Previous studies identified potential associated structures, with \citet{Landecker1989MNRAS} finding an H\MakeUppercase{\romannumeral1} cloud with a systemic velocity of $-34\pm 5$ $\kms$. More recently, \citet{Arias2019AA,Arias2024A&A} identified an H\MakeUppercase{\romannumeral1} cavity around $-$6 $\kms$ and a CO velocity gradient within [$-$6.5, 6.5] $\kms$. Furthermore, \citet{Huang1986ApJ} suggested that an MC with a CO velocity of approximately $-$22 $\kms$ may be spatially located near the remnant, suggesting a possible association. Taken together, these various kinematic results suggest a complex environment but have not yielded a consensus on the specific velocity component for the SNR-MC interaction.

\begin{figure}[ht!]
\centering
\includegraphics[width=0.7\textwidth]{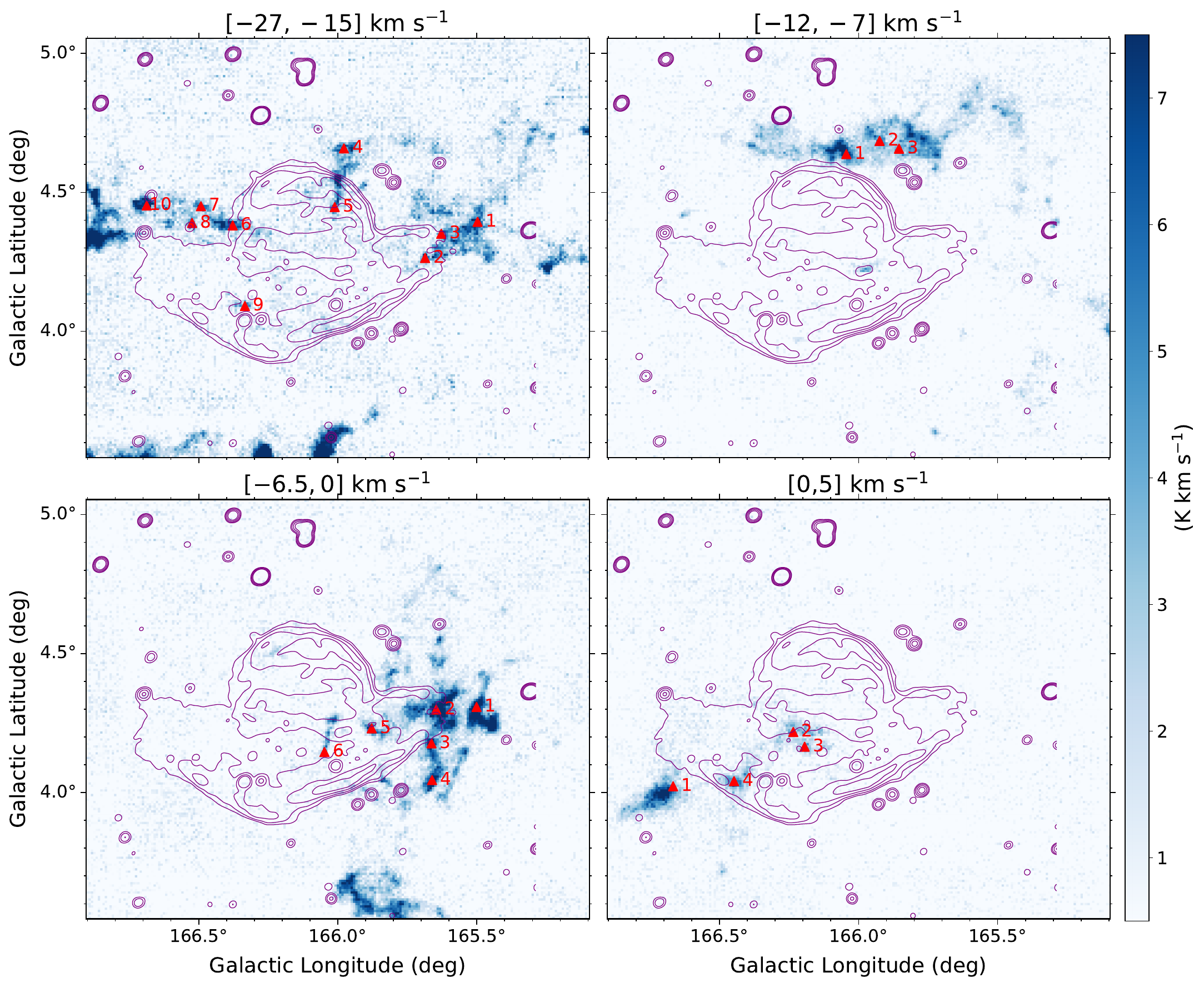}
\caption{Integrated intensity maps of $\CO$ for G166.0+4.3 covering velocity ranges of [$-$27, $-$15] $\kms$, [$-$12, $-$7] $\kms$, [$-$6.5, 0] $\kms$, and [0, 5] $\kms$ respectively (see Figure \ref{fig1}(d)). Purple contours represent the 1420 MHz radio emission shell of SNR, with the levels of 4.8, 5.0, 5.4, and 6.0 K. The red triangles are used for the same purpose as in Figure \ref{fig2}.} 
\label{fig11}  
\end{figure}

Utilizing high resolution MWISP $^{12}$CO observations, we have revealed four distinct velocity components in the molecular environment toward G166.0+4.3 in Figure \ref{fig1}(d): $[-27, -15]$ $\kms$, $[-12, -7]$ $\kms$, $[-6.5, 0]$ $\kms$, and [0, 5] $\kms$. The newly revealed velocity components enable a more precise assessment of previous kinematic suggestions. Specifically, the peak velocity ($\sim -21$ $\kms$) of the $[-27, -15]$ $\kms$ component is highly consistent with the $-$22 $\kms$ of the MC suggested by \citet{Huang1986ApJ}. Similarly, the $[-6.5, 0]$ $\kms$ component (peak velocity $\sim -$3.7 $\kms$) is kinematically related to the $-$6 $\kms$ H\MakeUppercase{\romannumeral1} structures found by \citet{Arias2019AA,Arias2024A&A}, though a slight discrepancy in peak velocity exists due to different data sources. Morphologically, the $[-6.5, 0]$ $\kms$ component is distributed along the western edge of SNR, while the $[-27, -15]$ component consists of three smaller MCs located in the eastern, western, and northern parts respectively in Figure \ref{fig11}. The spatial adjacency of both component boundaries to the radio shell strengthens the likelihood of association and necessitates further investigation into their distance. Based on these agreements and their spatial arrangement relative to the remnant, we propose that the $[-27, -15]$ $\kms$ and $[-6.5, 0]$ $\kms$ components are the most promising candidates for SNR-MC interaction. 

\begin{figure}[ht!]
\centering
\includegraphics[width=0.75\textwidth]{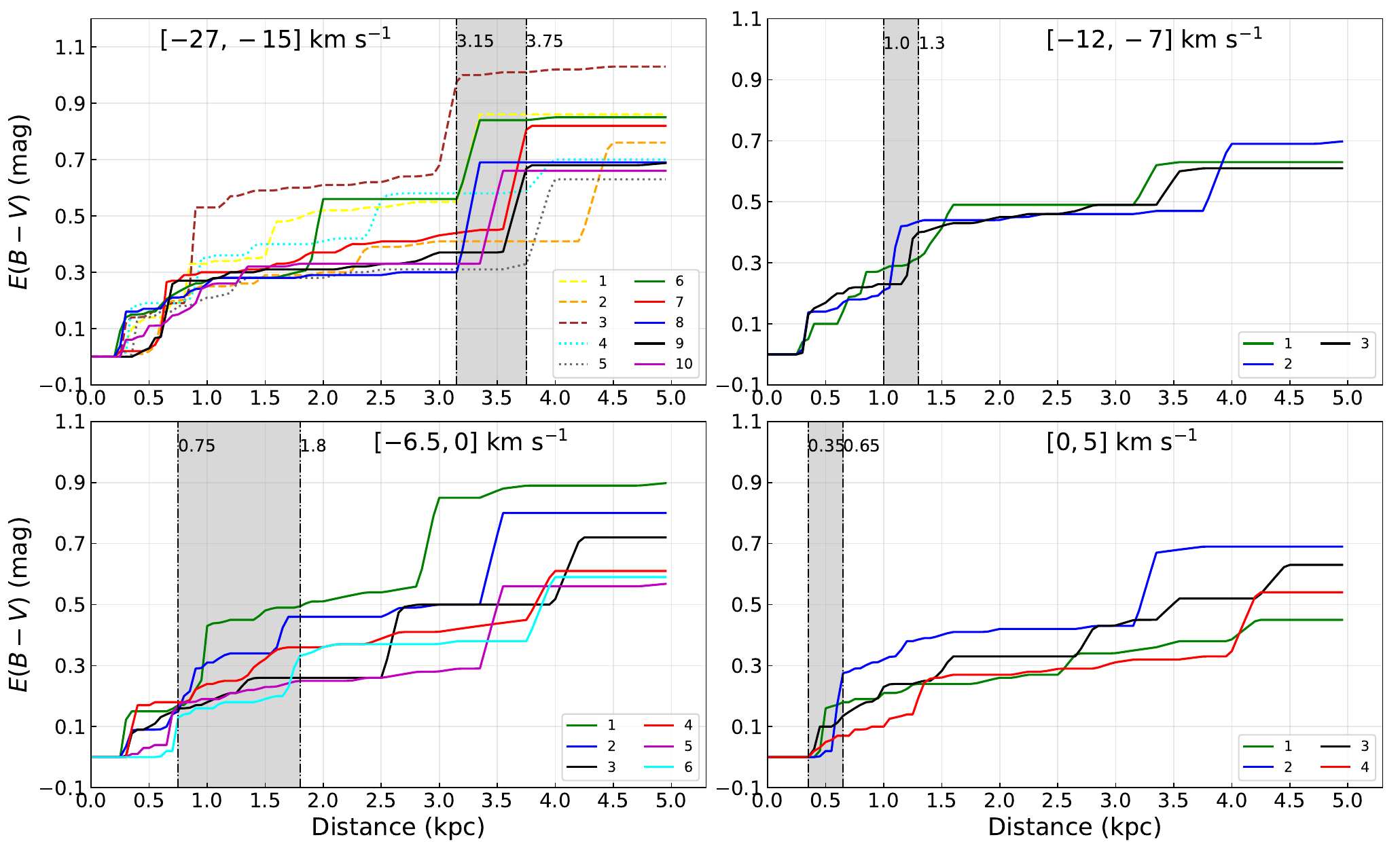}
\caption{The color excess $\EBV$ as a function of distance for G166.0+4.3 derived from 3D extinction map. Different colored lines correspond to the red positions marked in Figure \ref{fig11}. Vertical gray bands indicate extinction jumps for the corresponding MC: 3.15--3.75 kpc for $[-27, -15]$ $\kms$, 0.35--0.65 kpc for [0, 5] $\kms$, and a broad range of 0.75--1.8 kpc for both $[-12, -7]$ $\kms$ and $[-6.5, 0]$ $\kms$. For the $[-27, -15]$ $\kms$ component, the solid, dashed, and dotted lines represent triangles derived from three different molecular clumps shown in Figure \ref{fig11}.}
\label{fig12}
\end{figure}

\begin{figure}[ht!]
\centering
\includegraphics[width=0.8\textwidth]{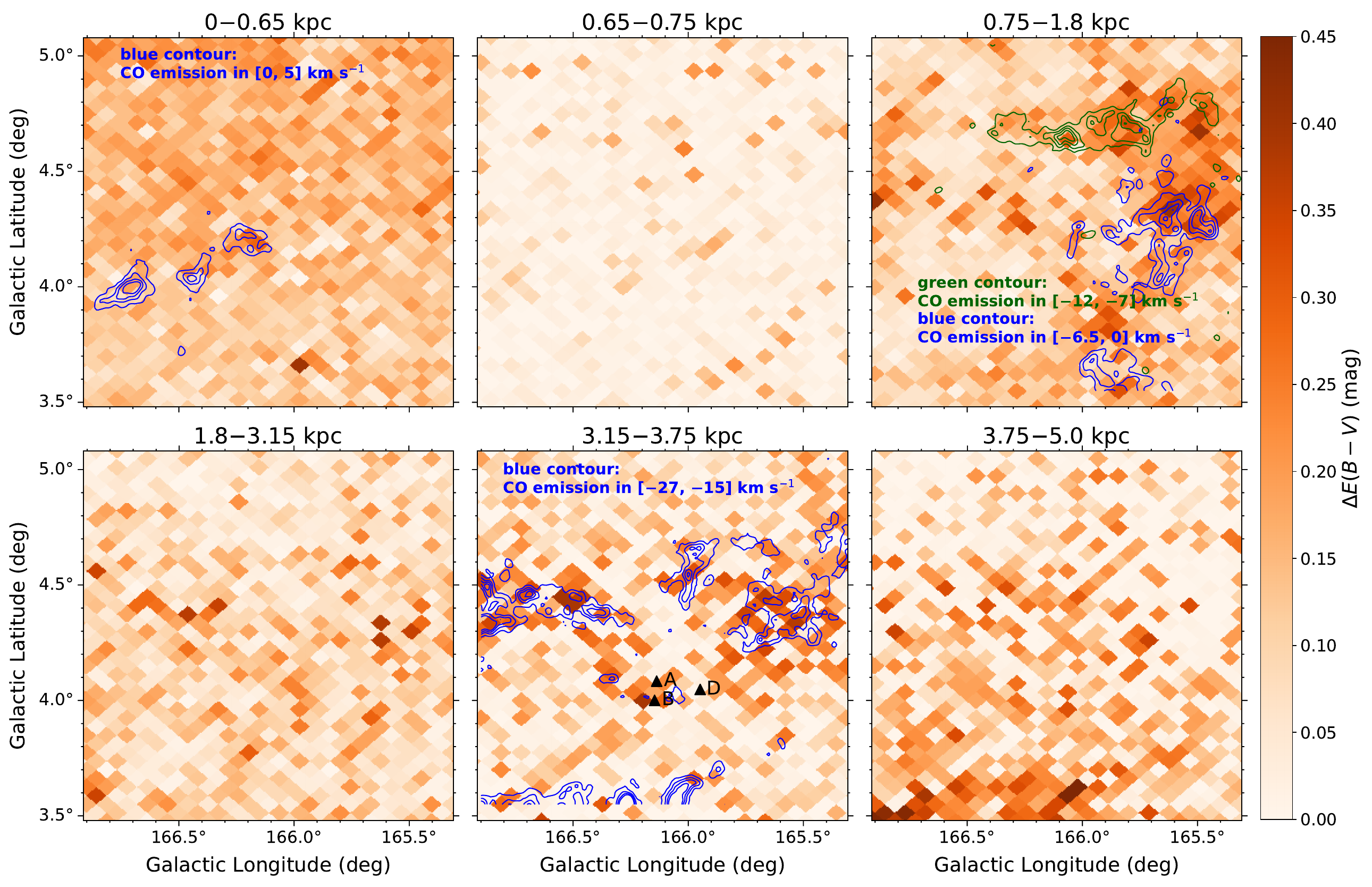}
\caption{Differential extinction maps ($\Delta E(B-V)$) toward G166.0+4.3 for distance bins of 0--0.65 kpc, 0.65--0.75 kpc, 0.75--1.8 kpc, 1.8--3.15 kpc, 3.15--3.75 kpc, and 3.75--5.0 kpc respectively. Blue contours show $\CO$ emission for the [0, 5] $\kms$ (0--0.65 kpc bin) and $[-27, -15]$ $\kms$ (3.15--3.75 kpc bin) components. The 0.75--1.8 kpc bin includes contours for the $[-12, -7]$ $\kms$ (green) and $[-6.5, 0]$ $\kms$ (blue) components. Contours are plotted at levels of 1.5, 3.5, 5.5, and 7.5 $\kkms$ for $[-27, -15]$ $\kms$ component; 1.6, 3.2, 4.8, and 6.4 $\kkms$ for $[-12, -7]$ $\kms$ component; 2.0, 4.5, 7.0, and 9.5 $\kkms$ for $[-6.5, 0]$ $\kms$ component; 1.5, 3.0, 4.5, and 6.0 $\kkms$ for [0, 5] $\kms$ component.
Black triangles labeled A, B, and D mark SNR-MC interaction regions identified by \citet{Arias2024A&A}. The color bar indicates $\Delta E(B-V)$ values ranging from 0 to 0.45 mag.}
\label{fig13}
\end{figure}

\begin{figure}[ht!]
\centering
\includegraphics[width=0.8\textwidth]{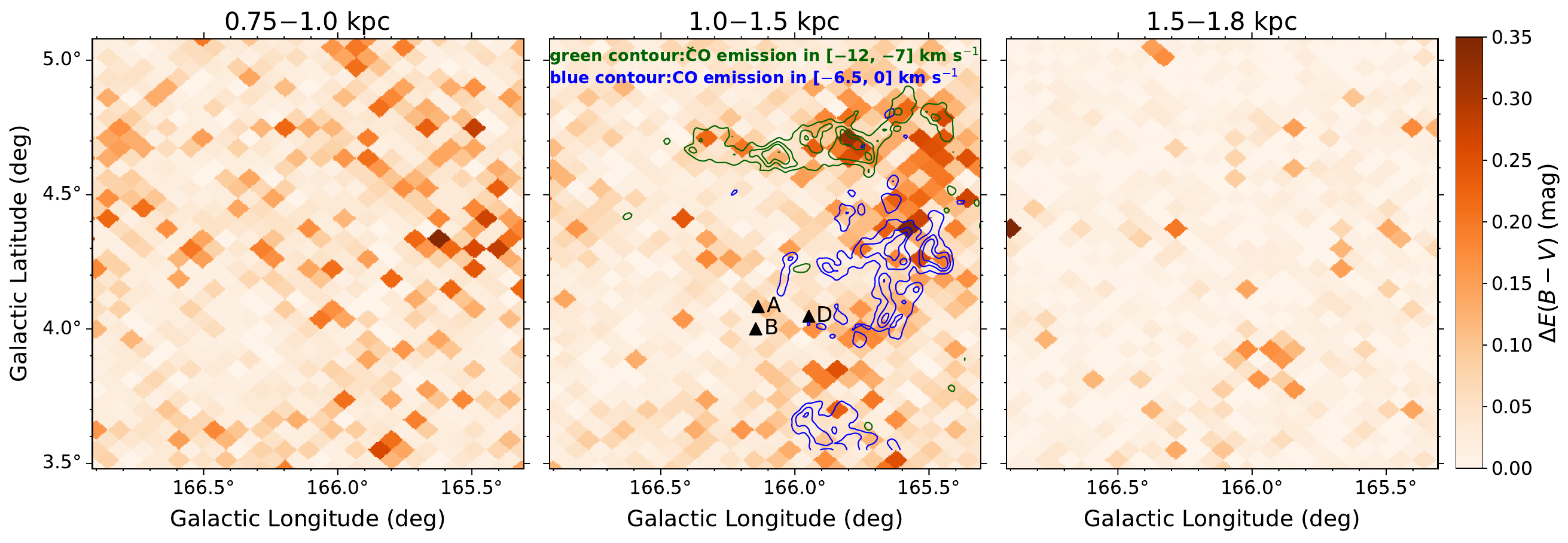}
\caption{Detailed differential extinction maps ($\Delta E(B-V)$) for G166.0+4.3 within the 0.75--1.8 kpc range, divided into distance sub-bins of 0.75--1.0 kpc, 1.0--1.5 kpc, and 1.5--1.8 kpc respectively. The 1.0--1.5 kpc bin is overlaid with CO emission contours for two molecular clumps: green for the [$-$12, $-$7] $\kms$ and blue for the [$-$6.5, 0] $\kms$. The contour levels for [$-$12, $-$7] $\kms$ and [$-$6.5, 0] $\kms$ are the same as those adopted in Figure \ref{fig13}. Black triangles labeled A, B, and D mark SNR-MC interaction regions identified by \citet{Arias2024A&A}. The color bar indicates the $\Delta E(B - V)$ value ranging from 0 to 0.35 mag.}
\label{fig14}
\end{figure}

To accurately constrain the distances of two proposed interacting components, $[-6.5, 0]$ $\kms$ and $[-27, -15]$ $\kms$, we constructed extinction-distance profiles in Figure \ref{fig12}. We first note that the other components show concentrated jumps at different distances: $[0, 5]$ $\kms$ is located at $\sim 0.5$ kpc, and $[-12, -7]$ $\kms$ is at $\sim 1.2$ kpc. Turning to our primary candidates, the $[-6.5, 0]$ $\kms$ component shows dispersed increases over a broad distance range of 0.75--1.8 kpc, suggesting it spans a large extent along the sightline and lies mostly in front of the high-velocity gas. In contrast, the $[-27, -15]$ $\kms$ component exhibits a prominent jump occurs at $\sim 3.5$ kpc. Therefore, our analysis focused on the broad 0.75--1.8 kpc range for $[-6.5, 0]$ $\kms$ component and the 3.15--3.75 kpc range for $[-27, -15]$ $\kms$ component. 
Based on these extinction jumps, we divide the extinction maps into six distance bins: 0--0.65 kpc, 0.65--0.75 kpc, 0.75--1.8 kpc, 1.8--3.15 kpc, 3.15--3.75 kpc, and 3.75--5.0 kpc, as shown in Figure \ref{fig13}.
Upon detailed sub-binning in Figure \ref{fig14}, we found the $[-6.5, 0]$ $\kms$ component is spatially located within the 1.0--1.5 kpc bin, a result consistent with the distance of $1.0\pm0.4$ kpc estimated by \citet{Arias2024A&A}. However, when testing the specific regions A, B, and D previously identified by \citet{Arias2024A&A} as evidence of SNR-MC interaction, we found they do not spatially correspond to molecular clumps at the 1.0--1.5 kpc bin in MWISP CO data. Instead, these key interaction regions exhibit significant extinction in the much farther 3.15--3.75 kpc bin (as marked in Figure \ref{fig13}). Furthermore, the extinction map in this 3.15--3.75 kpc bin shows excellent morphological correspondence with the $[-27, -15]$ $\kms$ component. Therefore, we suggest that the distance bin of 3.15--3.75 kpc, associated with the $[-27, -15]$ $\kms$ component, provides the more robust location for the MCs interacting with the SNR. 

Based on the confirmed association between the $[-27, -15]$ $\kms$ component and the remnant, the distance to G166.0+4.3 is constrained to approximately 3.4 kpc. This result is significantly larger than the $1.0\pm0.4$ kpc estimate derived from H\MakeUppercase{\romannumeral1} observations by \citet{Arias2019AA}, but smaller than the H\MakeUppercase{\romannumeral1}-based kinematic distance of $4.5\pm1.5$ kpc estimated by \citet{Landecker1989MNRAS}. Importantly, our distance is in good agreement with the recent extinction distance of $3.24\pm0.03$ kpc estimated by \citet{Zhao2020ApJ}. All results are summarized in Table \ref{table5}. Based on the robust morphological correlation between the radio shell and the $[-27, -15]$ $\kms$ component within the 3D extinction map, we estimate that a distance of $3.44 \pm 0.23$ kpc is the most reliable determination for G166.0+4.3.

\section{Discussion} \label{sec: discu}

\subsection{Application of method in the inner Galaxy}
In the inner Galaxy, kinematic distances derived from radial velocities suffer from the Kinematic Distance Ambiguity (KDA), where a single velocity corresponds to two possible distances (near and far). A major advantage of our method, which combines the 3D extinction map of \citet{Green2019ApJ} with molecular line observations, is its potential to resolve this ambiguity. Since the 3D extinction distance is determined independently of Galactic rotation, we can avoid the KDA by checking where the extinction jump occurs along the sightline. If the sharp rise in extinction matches the near kinematic distance, the KDA is resolved in favor of the near solution. However, the application of this method in the inner Galaxy has limitations due to severe dust obscuration. In extremely high extinction regions, the background stars used to construct 3D dust maps may become too faint to be detected, leading to large uncertainties or a lack of data at larger distances. Therefore, while our method is powerful for resolving KDA, its reliability in the inner Galaxy is limited by the maximum depth and sensitivity of the 3D extinction maps.

\subsection{$\gamma$-ray emission as evidence for SNR-MC interaction}
In addition to kinematic evidence from molecular lines, other indicators such as shocked gas tracers and $\gamma$-ray emission provide critical support for SNR-MC interactions. In particular, the $\gamma$-ray observations serve as direct evidence for the interaction between accelerated cosmic rays and dense molecular material. The decay of neutral pions produced in proton-proton collisions typically results in GeV $\gamma$-ray emission. Among the SNRs in our sample, G109.1$-$1.0 has been detected by the Fermi Large Area Telescope (LAT) \citep{Castro2012ApJ}, strongly supporting its association with MC. We checked the Fermi-LAT SNR catalog of \citet{Acero2016ApJS} for the other three SNRs in our sample, but they are listed as undetected. However, the lack of $\gamma$-ray detection does not necessarily rule out interaction, as it may be due to the limited sensitivity of current instruments or environmental factors affecting cosmic ray acceleration and diffusion.

\begin{figure}[ht!]
\centering
\includegraphics[width=0.9\textwidth]{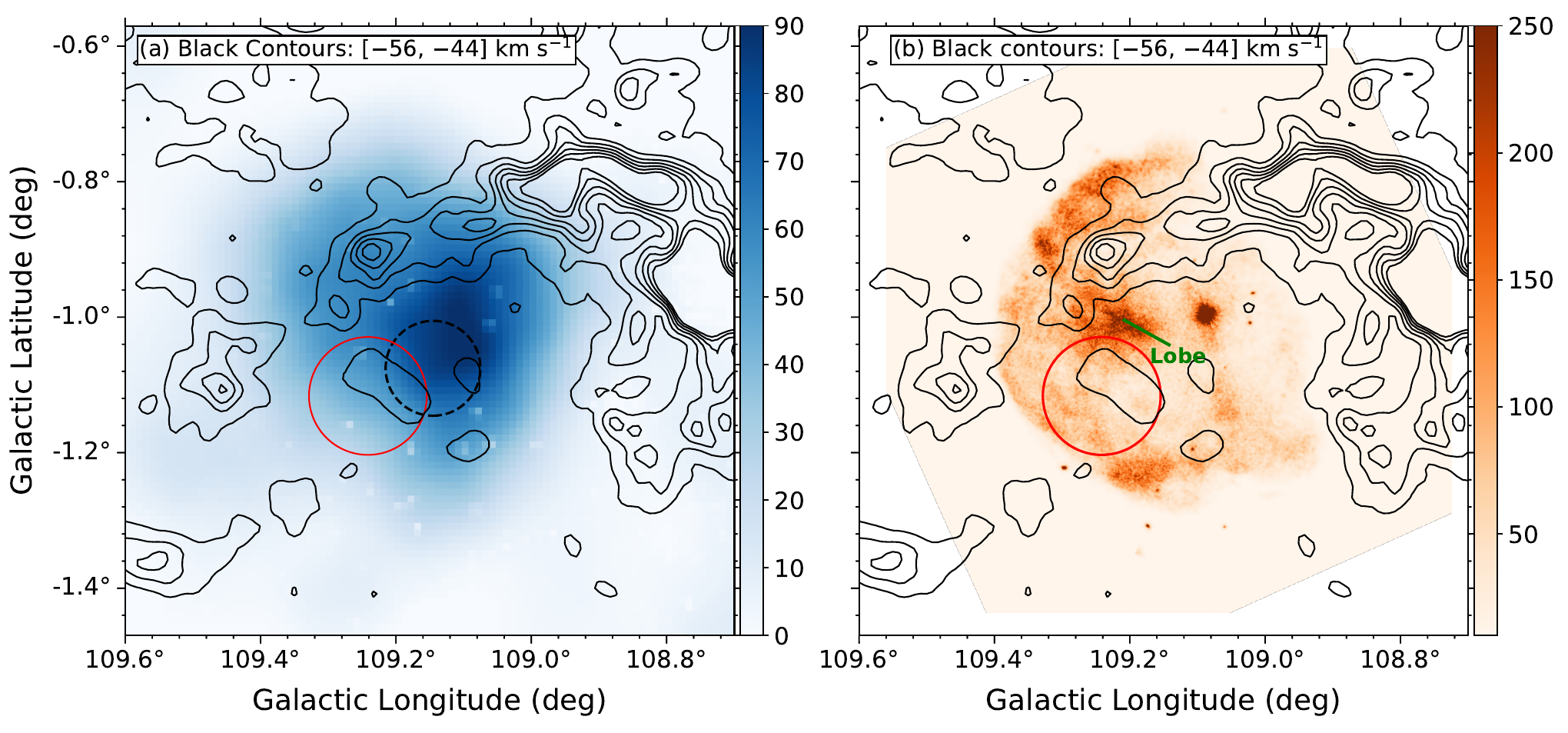}
\caption{The test statistic (TS) map centered at G109.1$-$1.0 for photons above 10 GeV from \citet{Xin2023ApJ} (a) and Intensity map (0.4--1.3 keV) of G109.1$-$1.0 from XMM-Newton data (b). The red solid circles denote the molecular clump in the 2.5--2.8 kpc bin in Figure \ref{fig7}. The black dashed circle marks the centroid position as a point source with its 95$\per$ confidence radius in \citet{Castro2012ApJ}. The two panels include the same contours of the $[-56, -44]$ $\kms$ component. Contours are plotted with the levels from 3.0 to 45.0 $\kkms$ with an interval of 7.0 $\kkms$. The strong X-ray emission region in the SNR center in panel (b) is the Lobe in \citet{Sasaki2004ApJ}.
}
\label{fig15}
\end{figure}

For G109.1$-$1.0, a spatial offset between the peak of CO emission and $\gamma$-ray peak is evident, as shown in Figure \ref{fig15} (a). The red circles mark the molecular clump identified in the 2.5--2.8 kpc bin in Figure \ref{fig7}. The $[-56, -44]$ $\kms$ component includes both the CO ridge and GMC ($[-51, -46]$ $\kms$), as well as several smaller molecular clumps located on the eastern part ($[-56, -51]$ $\kms$).
The $\gamma$-ray emission exhibits a centrally brightened morphology, which is inconsistent with the  cloud complex observed in the CO emission. Notably, the $\gamma$-ray centroid lies between the CO ridge and the molecular clump. This positional discrepancy may have multiple explanations. On one hand, the peak of CO intensity map traces a large amount of undisturbed molecular gas (i.e., the main body of cloud) \citep{Kothes2002ApJ}. In contrast, the $\gamma$-ray peak represents the location of shock front, where intense particle collisions occur \citep{Castro2012ApJ}. If the shock encounters the MC from the side, the interaction may occur at the cloud's boundary, naturally producing a spatial offset.
On other hand, \citet{Xin2023ApJ} suggested that the spatial offset could be attributed to dissociation of molecules by the SNR radiative precursor due to the photoionization and photodissociation effects, or the absence of CO emission may stem from the CO dark gas in the $\gamma$-ray emitting region that cannot be traced by CO observations.

In addition, \citet{Castro2012ApJ} proposed that it is possible that the Fermi-LAT source is related to an interaction between the SNR and the shocked molecular material in the X-ray ``Lobe" region in \citet{Sasaki2004ApJ}. This association was later confirmed by \citet{Sasaki2013A&A}. 
These studies consistently point to an interaction between G109.1$-$1.0 and its eastern Lobe, as illustrated in Figure \ref{fig15} (b). The molecular clump in the 2.5--2.8 kpc bin is located south of Lobe, suggesting a possible physical association. However, \citet{Tu2025} reported no evidence of line broadening in the Lobe region based on HCO$^{+}$ observations. Instead, they detected blue-shifted line wings north of SNR (tip of CO ridge), which may be due to the shock interaction. This result implies that the shock of G109.1$-$1.0 may be interacting simultaneously with both the CO ridge and eastern molecular clump. 
Overall, we suggest that the identification of an SNR-MC interaction associated with the western GMC remains reasonable.

\subsection{Limitations caused by other reasons}
Despite its strengths, our method has a few inherent limitations. Firstly, not all molecular gas is traced by CO emission, and diffuse or low-density clouds might be missed, potentially leading to an incomplete identification of interacting material. Then, the current spatial resolution and distance uncertainties of the 3D extinction map restrict the precision and applicability of the technique, especially for the MCs interacting with SNRs in small angular sizes or those located at very large distances where stellar tracers become rare.

In addition, it may be much harder to recognize interaction based solely on morphology if the SNR interacts with an MC on one side. \citet{Ranasinghe2017ApJ} pointed out that an MC coincidentally along the sightline to SNR in most cases does not show a clear correlation with those of the SNR border or contours. 
However, there are also instances where an MC is morphologically coincidental with an SNR without any compelling evidence for a physical interaction. Therefore, in such cases, we incorporate evidence with broadened CO line profiles from previous literature as supportive indicators for SNR-MC interaction.

Future improvements will stem from methodological advancements and new observational data. 
Incorporating additional tracers of shocked molecular gas could provide more robust confirmation of SNR-MC interactions. For example, \citet{Rho2021ApJ} provided clear evidence for the interaction between HB 3 and an MC through the detection of shock-excited H$_{2}$ emission and  CO broad lines, and further revealed the large-scale structure of the interaction region. \citet{Lee2020AJ} conducted near-infrared spectroscopic observations of 16 SNRs exhibiting strong H$_{2}$ emission lines and demonstrated that these features are most likely associated with SNR-MC interactions. \citet{Tu2025} identified evidence for SNR-MC interactions in eight remnants based on HCO$^{+}$ line emission.

The ongoing development of next-generation, higher-precision 3D extinction maps, leveraging data from Gaia DR3/DR4 and deeper multi-band photometric surveys, will significantly enhance the distance accuracy and spatial detail available for this analysis. Such advancements will extend the applicability of this powerful combined technique to a broader range of SNRs and further refine our understanding of their Galactic distribution and evolution.


\section{Summary} \label{sec: sum}

We present a novel method for determining distances to SNRs by combining CO line observations with 3D dust extinction map. Our two-step approach first identifies MCs physically associated with an SNR through analysis of CO spectra and the spatial correspondence between CO emission and the SNR's radio shell with the help of literature. Second, the extinction-distance profiles toward these candidate MCs are extracted from the 3D extinction map, from which the significant extinction jumps are selected, and the morphology of differential extinction maps within specific distance bins are compared with the CO velocity-integrated intensity maps. The distance at which the morphological correspondence is strongest is assigned to the MC and to the SNR itself by the physical association. This dual-constraint methodology effectively minimizes contamination from unrelated foreground or background clouds along the sightline.

We applied this method to four SNRs: G93.7$-$0.2, G109.1$-$1.0, G156.2$+$5.7, and G166.0$+$4.3. Our analysis yields distances of $1.82 \pm 0.13$ kpc for G93.7$-$0.2 associated with the CO  \textit{v} $\in$ [$-$19, $-$3] $\kms$ component; $3.05 \pm 0.15$ kpc for G109.1$-$1.0 associated with the CO \textit{v} $\in$ [$-$51, $-$46] $\kms$ component; $0.60 \pm 0.15$ kpc for G156.2+5.7 associated with the CO \textit{v} $\in$ [$-$10, 0] $\kms$ component; and $3.44 \pm 0.23$ kpc for G166.0+4.3 associated with the CO \textit{v} $\in$ [$-$27, $-$15] $\kms$ component. 
These results are largely consistent with previous independent estimates and with better certainty. The method proves particularly effective in complex interstellar environments where multiple MCs exist along the sightline, as it can isolate the specific cloud interacting with the remnant based on kinematic and morphological coherence.

\begin{acknowledgments}
We are grateful to Drs. Shu Wang and Jian Gao for their helpful discussion and suggestion. This work is supported by the National SKA Program of China No. 2025SKA0140100, and the National Natural Science Foundation of China (NSFC) project Nos. 12133002 and 12403026. This research made use of the data from the Milky Way Imaging Scroll Painting (MWISP) project, which is a multi-line survey in $\CO/\COO/\COOO$ along the northern galactic plane with PMO-13.7m telescope. We are grateful to all the members of the MWISP working group, particularly the staff members at PMO-13.7m telescope, for their long-term support. MWISP was sponsored by National Key R$\&$D Program of China with grant 2017YFA0402701 and CAS Key Research Program of Frontier Sciences with grant QYZDJ-SSW-SLH047.

\end{acknowledgments}

\vspace{5mm}
\facilities{PMO-13.7m, CfA-1.2m}

\software{Astropy \citep{AstropyCollaboration2013A&A},
          APLpy \citep{Robitaille2012ascl.soft},
          dustmaps \citep{Green2018JOSS},
          Spectral-cube \citep{Ginsburg2015ASPC}
          }

\bibliography{sample701}{}
\bibliographystyle{aasjournalv7}

\begin{table}[h]
\centering
\renewcommand{\arraystretch}{1.2}
\caption{The regions, velocity components, and other velocity evidences for all the four SNRs}
\label{table1}
\resizebox{\textwidth}{!}{
\begin{tabular}{c c c c c c}
\hline \hline
Name & Size & Aperture radius & Velocity components (this work) & other evidence$^{(a)}$ & References\\
  & (arcmin) & (arcmin) & ($\kms$) & ($\kms$) & \\
\hline
G93.7$-$0.2 & 80 & 48 & [$-$55, $-$30], [$-$19, $-$3], [$-$3, 6], [7, 15]& $-$6 (H\MakeUppercase{\romannumeral1}) & 1\\
G109.1$-$1.0 & 28 & 19.8 & [$-$56, $-$44], [$-$12, $-$6.5] & [$-$57, $-$43] (CO),$-$56 (H\MakeUppercase{\romannumeral1}),[$-$52, $-$47] (CO), $-$51 (CO) & 2--6 \\
G156.2+5.7 & 110 & 63 & [$-$10, 0], [0, 10]& $[-8, 8]^{(b)}$ (CO), [$-$48, $-$41] (H\MakeUppercase{\romannumeral1})& 7\\
G166.0+4.3 &  $55 \times 35$ & $40.5 \times 25.5^{(c)}$ & [$-$27, $-$15], [$-$12, $-$7], [$-$6.5, 0], [0, 5]& $-$22 (CO), $-$34 (H\MakeUppercase{\romannumeral1}), $-$6 (H\MakeUppercase{\romannumeral1}) & 8--10\\
\hline
\end{tabular}}
\tablecomments{\scriptsize 
(a) $\vlsr$ of spatially correlated CO emission and H\MakeUppercase{\romannumeral1} observations.\\
(b) \citet{Reich1992AA} detected CO emission in range of [$-$8, 8] $\kms$ in the SNR region, but argued that the corresponding MC places in front of the remnant.\\
(c) Since the aperture shape is an ellipse, the values here are the semi-major and semi-minor axes.\\
\textbf{References:}
(1) \citet{Uyaniker2002ApJ}, 
(2) \citet{Tatematsu1990ApJ}, (3) \citet{Sasaki2006ApJ}, (4) \citet{Tian2010MNRAS}, (5) \citet{Kothes2002ApJ}, (6) \citet{Zhou2023ApJS},
(7) \citet{Reich1992AA}, 
(8) \citet{Huang1986ApJ}, (9) \citet{Landecker1989MNRAS}, (10) \citet{Arias2019AA}
}
\end{table}

\begin{table}[h]
\centering
\caption{The distances to the G93.7$-$0.2}
\label{table2}
\begin{tabular}{c c c}
\hline \hline
References & Methods & Distance \\
   &  & (\kpc) \\
\hline
\citet{Ranasinghe2022ApJ} & kinematics & 1.3$\pm$0.2 \\
\citet{Uyaniker2002ApJ} & kinematics & 1.5$\pm$0.2 \\
\citet{Wang2020AA} & RCs & 1.99$\pm$0.03 \\ 
\citet{Zhao2020ApJ} & extinction & 2.16$\pm$0.02 \\
This work & 3D extinction & $1.82 \pm 0.13$ \\

\hline
\end{tabular}
\end{table}

\begin{table}[h]
\centering
\caption{The distances to the G109.1$-$1.0}
\label{table3}
\begin{tabular}{c c c}
\hline \hline
References & Methods & Distance \\
   &  & (\kpc) \\
\hline
\citet{Zhao2020ApJ} & extinction & 2.79$\pm$0.04 \\
\citet{Kothes2002ApJ} & kinematics & 3.0$\pm$0.5 \\
\citet{Sanchez2018MNRAS} & kinematics & 3.1$\pm$0.2 \\ 
\citet{Chen2025ApJ} & extinction & $\sim$3.12 \\
\citet{Tian2010MNRAS} & kinematics & 4.0$\pm$0.8 \\
\citet{Durant2006ApJ} & RCs & 7.5$\pm$1.0 \\
This work & 3D extinction & $3.05 \pm 0.15$ \\

\hline
\end{tabular}
\end{table}

\begin{table}[h]
\centering
\caption{The distances to the G156.2+5.7}
\label{table4}
\begin{tabular}{c c c}
\hline \hline
References & Methods & Distance \\
   &  & (\kpc) \\
\hline
\citet{Gerardy2007MNRAS} & associated object & 0.3--0.6 \\
\citet{Zhao2020ApJ} & extinction & 0.68$\pm$0.2 \\
\citet{Reich1992AA} & kinematics & 1--3 \\ 
\citet{Yamauchi1999PASJ} & Sedov estimate & 1.3 \\
\citet{Katsuda2016ApJ} & proper motion and shock velocity & $\textgreater$1.7 \\
This work & 3D extinction & $0.60 \pm 0.15$ \\

\hline
\end{tabular}
\end{table}

\begin{table}[h]
\centering
\caption{The distances to the G166.0+4.3}
\label{table5}
\begin{tabular}{c c c}
\hline \hline
References & Methods & Distance \\
   &  & (\kpc) \\
\hline
\citet{Arias2019AA} & kinematics & 1.0$\pm$0.4 \\
\citet{Zhao2020ApJ} & extinction & 3.24$\pm$0.03 \\
\citet{Landecker1989MNRAS} & kinematics & 4.5$\pm$1.5 \\ 
This work & 3D extinction & $3.44 \pm 0.23$ \\

\hline
\end{tabular}
\end{table}


\end{CJK*}
\end{document}